\documentclass[fleqn,usenatbib]{rasti}
\usepackage{newtxtext,newtxmath}

\usepackage[T1]{fontenc}
\usepackage{mathtools, cuted}
\usepackage{multirow}

\DeclareRobustCommand{\VAN}[3]{#2}
\let\VANthebibliography\thebibliography
\def\thebibliography{\DeclareRobustCommand{\VAN}[3]{##3}\VANthebibliography}

\usepackage{color}

\usepackage[version=4]{mhchem}
\usepackage{graphicx}
\usepackage{booktabs} 

\usepackage{graphicx}	
\usepackage{amsmath}	
\usepackage{subcaption}

\title[The impact of spectral line wing cut-off: Recommended standard method for the \texttt{MAESTRO} opacity database]{The impact of spectral line wing cut-off: Recommended standard method with application to {\tt MAESTRO} opacity database}

\author[Ehsan  Gharib-Nezhad et al.]{
Ehsan (Sam) Gharib-Nezhad,$^{1,2}$\thanks{Lead Author, e.gharibnezhad@asu.edu}\thanks{All authors have contributed equally to this study.}
Natasha E. Batalha,$^{1}$
Katy Chubb,$^{3}$
Richard Freedman,$^{1,4}$
Iouli E. Gordon,$^{5}$
\newauthor{
Robert R. Gamache,$^{6}$
Robert J. Hargreaves,$^{5}$
Nikole K. Lewis,$^{7}$
Jonathan Tennyson,$^{8}$
Sergei N. Yurchenko,$^{8}$}
\\
$^{1}$ Space Science and Astrobiology Division, NASA Ames Research Center, Moffett Field, CA, USA;\\
$^{2}$ Bay Area Environmental Research Institute, CA, USA;\\
$^{3}$ School of Physics and Astronomy, University of St Andrews, St Andrews, United Kingdom;\\
$^{4}$ Carl Sagan Center, SETI Institute, 515 North Whisman Road, Mountain
View, CA 94043.;\\
$^{5}$ Atomic and Molecular Physics, Harvard-Smithsonian Center for Astrophysics, Cambridge, MA, USA;\\
$^{6}$ Department of Environmental, Earth, and Atmospheric Sciences, University of Massachusetts, Lowell, MA, USA;\\
$^{7}$ Department of Astronomy, Cornell University, Ithaca, NY, USA;\\
$^{8}$ Department of Physics and Astronomy, University College London, London, United Kingdom;
}

\date{Accepted XXX. Received YYY; in original form ZZZ}

\pubyear{2023}

\begin{document}
\label{firstpage}
\pagerange{\pageref{firstpage}--\pageref{lastpage}}
\maketitle

\begin{abstract}
When computing  cross-sections from a line list, the result depends not only on the line strength, but also the line shape, pressure-broadening parameters, and line wing cut-off (i.e., the maximum distance calculated from each line centre). Pressure-broadening can be described using the Lorentz lineshape, but it is known to not represent the true absorption in the far wings. Both theory and experiment have shown that far from the line centre, non-Lorentzian behaviour controls the shape of the wings and the Lorentz lineshape fails to accurately characterize the absorption, leading to an underestimation or overestimation of the opacity continuum depending on the molecular species involved.  The line wing cut-off is an often overlooked parameter when calculating absorption cross sections, but can have a significant effect on the appearance of the spectrum since it dictates the extent of the line wing that contributes to the calculation either side of every line centre. Therefore, when used to analyse exoplanet and brown dwarf spectra, an inaccurate choice for the line wing cut-off can result in errors in the opacity continuum, which propagate into the modeled transit spectra, and ultimately impact/bias the interpretation of observational spectra, and the derived composition and thermal structure. Here, we examine the different methods commonly utilized to calculate the wing cut-off and propose a standard practice procedure (i.e., absolute value of 25~cm$^{-1}$ for $P\leqslant$~200~bar and 100~cm$^{-1}$ for $P >$ ~200~bar) to generate molecular opacities which will be used by the open-access {\tt MAESTRO} (Molecules and Atoms in Exoplanet Science: Tools and Resources for Opacities) database. The pressing need for  new measurements and theoretical studies of the far-wings is highlighted.

\end{abstract}




\section{Introduction}

Observations of a diverse set of atmospheres, from  brown dwarfs and hot gas giants, to small rocky worlds, will be a legacy of JWST and the future next-generation ground based telescopes. In order to robustly compare observations to models,  the community needs access to validated and up-to-date opacities. In many cases, computing opacities for a single molecule, under a specified pressure-temperature-gas mixture combination, is not trivial. For exoplanet conditions specifically (i.e., high temperature and exotic, non-air gas mixtures), data are often incomplete \citep[as outlined in][]{fortney2019need} and could lead to noticeable errors in modeling transit spectra of exoplanets \citep[e.g.,][]{GharibNezhad2019steam,22NiWiGo} and brown dwarf atmospheres \citep[e.g.,][]{iyer2023sphinx}. Beyond line list data use, other modeling choices in opacity calculations affect the resultant calculation, including line profile, intensity cutoffs, and wing cut-offs. This was the motivation for the {\tt MAESTRO}\footnote{\href{https://science.data.nasa.gov/opacities/}{science.data.nasa.gov/opacities/}} (Molecules and Atoms in Exoplanet Science: Tools and Resources for Opacities) database: an open NASA-supported opacity service that can be accessed by the community via a web interface. Of priority for the {\tt MAESTRO} project is to develop community standards in computing and publishing molecular and atomic opacity data. The main principles of the {\tt MAESTRO} interface are to highlight: 1) known updates from line lists communities that need to be incorporated into the community database, 2) opacity calculation cautions that specify when approximations or estimations have been made, 3) limits of use for comparing models to observations (e.g., completeness and line position accuracy), and 4) the full breadth of data (including relevant citations) that are used for a single opacity calculation. Currently, the focus of this work is on molecular, not atomic opacities. Atomic opacities will be the focus of future work. For the molecular opacity calculation cautions listed, we currently employ a set of standard warnings. We have six standard warnings that pertain to the broadening choices (e.g., ``Laboratory or theoretical broadener information not available at all'' or ``Broadener data for an air mixture was used in place of this broadener gas''), and two standard warnings that pertain to the line list choices (``Line list used is not suitable for this temperature regime'' and ``The line position accuracy is suitable for JWST-quality spectra only (R=100-3000)'')..

We aim to accept community contributions to the {\tt MAESTRO} opacity database, which will undergo an open peer review through our contributor portal on Github\footnote{\href{https://github.com/maestro-opacities/submit-data}{github.com/maestro-opacities/submit-data}}. In order to facilitate community contributions we must first agree on a standard method for wing cut-offs such that we can effectively execute community model comparisons. Therefore, here, we document our standard methods that we will adopt in the {\tt MAESTRO} database. This manuscript specifically focuses on the impact of different line wing cut-off methods and the ``best'' method implemented to generate opacity datasets using Voigt profile. In a future MAESTRO update, we aim to expand our calculation cautions to include those that pertain specifically to wing-cutoffs choices such that users can assess how choices in wing-cutoff might affect their opacity data.

The Voigt profile, which is a convolution of Lorentz and Doppler (Gaussian) profiles, is a common profile when generating model spectra  from molecular/atomic line lists. While this profile is extensively utilized for generating absorption cross sections (ACS), spectroscopic measurements have shown that non-Lorentzian behaviour plays a key role in both the line centre and the wings (For example, for the distortion from the line centre see \cite{13NgLiTr,16WcGoTr.H2,17NgLiHo,burch1963total, burch1971absorption, burch1968absorption, burch1967absorption, burch1956infrared, burch1962total}, and for the wings see \citep{Hartmann2002}). Establishing a consistent policy for determining the limit of the extent of the line profile is a difficult and complex problem. The far wing is usually defined as the region beyond a certain multiple of the line widths from the line centre. For most species, knowledge of the far wing may not be well established either by theory, experiment, or a combination of the two. This deviation from the perfect Lorentzian behaviour begins at $\sim$5 - 30 cm$^{-1}$ from the line centre, and depends on the absorber, broadening agent, pressure, and temperature.

The inclusion of the non-Lorentz behaviour in the ACS data is challenging due to the lack of complete spectroscopic parameters. The main technique used in astrophysics to minimize the issue is to introduce a wing cut-off, $R_{\rm cut}$, parameter. $R_{\rm cut}$ can be a fixed spectral distance (e.g., 25\,cm$^{-1}$ is typically used for calculations of terrestrial radiative transfer \citep{mlawer2012development}) applied to all lines in the ACS calculation equally, or it can be dependant on a property of each line. For example, some studies have used $R_{\rm cut} = 200 \gamma_{\mathrm{v}}$ to 500$\gamma_{\mathrm{v}}$ \citep{14FrLuFo,15GrHe,16HeMaxx,20ChRoAl,jt819}, where $\gamma_{\mathrm{v}}$ is the Voigt half-width-at-half-maximum (HWHM). In addition, a maximum cutoff can be used to prevent the line wings from becoming too large at higher pressures~\citep{20ChRoAl}.
Some studies have employed a fixed value for all or different pressure ranges \citep{MacDonaldPhD,21GhIyLi,GharibNezhad2021Li, zhang2020platon}, or even use a pressure-dependent cutoff, such as $\min(25P; 100)$~cm$^{-1}$\footnote{To prevent the absorption from becoming unrealistically large, a pressure-dependent cutoff with maximum value of 100 cm$^{-1}$ is proposed.}, where $P$ is the total gas pressure in atmospheres,  and also apply normalization correction factors to ensure that the total strength of the profile is conserved~\citep{07ShBuxx}. 
In another study, \citep{letchworth2007rapid} have proposed different equations to control the truncation of the line wing in under different P-T conditions including the elimination of the weak lines and evaluating the resultant errors.

A few studies, however, calculated semi-empirical parameters or ``correction factors'' to improve the Voigt result.  Examples are limited to few spectroscopic measurements and for room temperature including \ce{CH4}-in-\ce{H2} for $P_{\ce{H2}}=10-200$ atm \citep{Hartmann2002}, and \ce{CH4}-in-\ce{CO2} for $P_{\ce{H2}}=3-25$ bar  \citep{Tran2022wing}.  Correction factors are often different for different bands of a molecule; for instance, the Spectral Mapping Atmospheric Radiative Transfer model~\citep{SMART}  derives correction factors and their functional implementations for different bands of CO$_2$ based on existing experimental studies and spectra of Venus. The spectral wings generated from all these different methods are not  necessarily consistent, and hence different ACS spectra are generated and used for atmospheric modeling. This hinders cross comparisons between studies, even when the source line lists used are identical. As a result of different choices in the wing cut-off, this  imposes  difficulties when comparing the resultant physical inferences from substellar spectra.

In addition to the inconsistency problem, inaccurate choice of $R_{\rm cut}$ results in many types of errors. First, a small value of $R_{\rm cut}$ results in an underestimation of the opacity continuum, particularly at high pressures ($\gtrsim$200~bar) where the Voigt (or Lorentzian) HWHM  can be a few tens of wavenumbers. This means that  the wing region should be extended. Second, having a large value for $R_{\rm cut}$ leads to an overestimation of the opacity continuum at low pressures ($\lesssim10^{-2}$~bar) where the Voigt HWHM is $\lesssim$~0.01cm$^{-1}$. 
Third, far wing effects are often 
included in continuum models~\citep{mlawer2012development,16ShCaDi}
to correct the overall opacity continuum (see Section~\ref{sec:far-wing-challenges} below) and it is important to match
 $R_{\rm cut}$ with the definition of these functions to avoid over/under counting.
Another related problem specific to low pressure calculations is that the line profile is sometimes narrower than  the size of the wavelength grid intervals. This can lead to  under-sampling, or no
sampling at all, and impact the accuracy of the ACS\footnote{To generate absorption cross sections, a line-by-line code typically iterates over all transitions in the line list. A wavelength grid vector is created before the loop to accumulate the opacity values from the Voigt profile for each generated line profile within the loop. Subsequently, these values are added to the wavelength grid vector. At very low pressures and very low frequencies, the line widths become extremely narrow, potentially some of these lines will not be added accurately to the grid if they are very narrower than the chosen grid spacing. Therefore, the key point here is that the wavelength grid spacing should be sufficiently smaller than the spectral line widths to include the opacity contributions from all the lines and minimize the loss of intensity in such cases.
}. Among the solutions are  moving the line centres to the nearest grid point \citep{07ShBuxx} or using the line profile  binning over the grid point instead of  sampling \citep{13HiYuTe}. 
These inaccuracies become crucial in spectral regions with a large number of spectral lines because they add up and make the overall opacity continuum stronger or weaker.

Our main objective in this study is to propose a ``standard practice'' method to control inaccuracies resulted by  calculations of the non-Voigt behaviour for the {\tt MAESTRO} opacity database. {\tt MAESTRO} spans a wide range of pressures and temperatures (i.e., 10$^{-6}$--3000~bar and 75--4000~K). The justification for such a wide range of pressures and temperatures in the context of atmospheric radiative transfer modeling is provided by \S\ref{sec:astrophysical-context}. Spectroscopic drawbacks in calculating the non-Voigt profile and its impacts on far wings are discussed in \S\ref{sec:far-wing-challenges}. \S\ref{sec:challenges} deals with other challenges: the challenges with the theoretical calculation of the thermochemical phase of the materials/gases in a very high-pressure regime are discussed in \S\ref{sec:high-pressure-challenge}, and the impact of the window regions in 
\S\ref{sec:spectral-windows}.
Two popular methods for choosing the $R_{\rm cut}$ to mitigate the effect of non-Voigt behaviour of the spectral lines are quantitatively analyzed and discussed for \ce{CH4}-in-air system in \S\ref{sec:Benchmark-CH4-cutoff}.
 These two scenarios are then tested  using the experimental spectrum of \ce{CO2}-in-\ce{N2} system in \S\ref{sec:lab_spectra_co2}. The accuracy of the super-lines  technique and its impact on the opacity continuum for the \ce{SO2}-in-\ce{H2} system is assessed in \S\ref{sec:superlines}. Our recommendations for a current default line wing cut-off are presented in \S\ref{sec:recommendation}, followed by instructions on how to control the line wing cut-off parameter in some example well-used codes in \S\ref{sec:codes}. We finish with our conclusions in \S\ref{sec:conclusions}.

\section{Astrophysical context for the temperature and pressure ranges }\label{sec:astrophysical-context}

\begin{figure*}

	\includegraphics[scale=0.66]{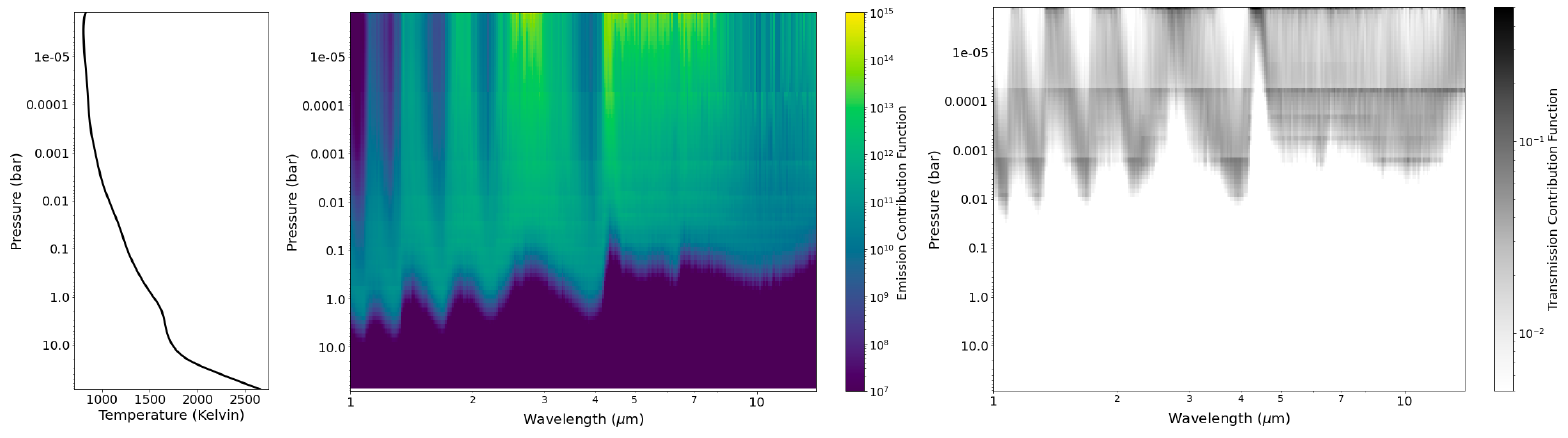}
    \caption{Typical hot Jupiter pressure-temperature profile (left) and associated contribution functions over relevant JWST wavelengths. Contribution functions dictate the pressure regions that the observable spectrum is sensitive to. The emission contribution function (middle) is defined in \citet{Lothringer2018ApJ}. The transmission contribution (right) is defined in \citet{molliere2019}. 
    }
    \label{fig:TPprofile-ContributionFunc}
\end{figure*}

Missions to the gas giant planets in our own solar system (e.g., Cassini, Galileo, Juno) have long sought to probe their deep atmospheres, which critically shape the chemistry and circulation patterns in the visible portion of their atmospheres. Since we cannot yet send spacecraft to visit planets orbiting other stars, we must rely on a combination of remote observations and atmospheric modeling to tie together the processes in their deep atmospheres/interiors and observable upper atmospheres. Theoretical models of giant exoplanet atmospheres critically depend on assumptions about atmospheric opacity sources which control the radiative balance between external and internal heating sources. Internal heat in giant exoplanets arising from both their formation and external forcing critically shapes the deep adiabatic region of their atmospheres (pressures greater than 10s to 100s of bars) which feeds into the radiative processes in the upper atmosphere \citep[e.g.][]{thorngren2019intrinsic}. Figure~\ref{fig:TPprofile-ContributionFunc} illustrates the thermal structure in an atmosphere of a generic hot Jupiter across the pressure-temperature space (left), and associated contribution functions over the relevant JWST  wavelengths for thermal emission (middle) \citep{Lothringer2018ApJ} and transmission  \citep{molliere2019} observations.

Additionally, giant planets orbiting close to their host stars are expected to have deep atmospheric radiative zones extending to pressures of 1000 bar or more \citep[e.g.][]{Showman2002,Fortney2005ApJ}. In order to attempt to capture the radiative, advective, and chemical processes that occur throughout giant exoplanet atmospheres, we must make assumptions about the atmospheric opacity at pressures well beyond the 1-10 bars typically considered. Hence, our opacity data for the {\tt MAESTRO} database are calculated for a wide range of pressures, 10$^{-6}$--3000~bar, to be applicable to modeling all cases. Table~\ref{tab:comp} lists the temperature and pressure grid points in detail.

\begin{table}
  \centering 
	\scriptsize   
  \caption{Opacity computational details for {\tt MAESTRO} database}\label{tab:comp}     
     \begin{tabular}{llllllllll}
       \toprule[\heavyrulewidth]\toprule[\heavyrulewidth]
     \multicolumn{7}{l}{\bf T[K] grid: (73 points)}\\
    &  75 & 100  & 110 & 120 & 130 &  140 & 150 & 160  \\
    & 170 &  180 &  190  &  200 & 210 &  220 & 230 & 240 \\     
    &  250&  260 & 270 &  275 & 280 & 290  & 300 &  310\\ 
     &  320 & 330&  340 & 350 &  375 & 400  & 425 &  450 \\ 
    &  475 & 500&  525 & 550 &  575 & 600  & 650 &  700 \\ 
    &  750 & 800&  850 & 900 & 950 &  1000 & 1100 & 1200\\ 
     & 1300 & 1400&  1500 &  1600 & 1700 & 1800 &  1900 &  2000 \\
      & 2100 & 2200 &  2300&  2400 & 2500 &  2600 &  2700 & 2800 \\
      & 2900 &  3000 &  3100& 3200 &  3300 &  3400 & 3500 & 3750 \\
    & 4000 \\
	\midrule[0.3pt]
     \multicolumn{7}{l}{\bf P[bar] grid: (20 points)}\\
 & 10$^{-6}$ & 3$\times$10$^{-6}$ & 10$^{-5}$ & 3$\times$10$^{-5}$ & 10$^{-4}$ & 3$\times$10$^{-4}$ \\
  & 10$^{-3}$ & 3$\times$10$^{-3}$ & 10$^{-1}$  & 3$\times$10$^{-1}$ & 10$^{-2}$ & 3$\times$10$^{-2}$    \\ 
 & 1 & 3  & 10 & 30  &  100 &   300  \\
 & 1000& 3000\\
    \bottomrule[\heavyrulewidth] 
   \end{tabular}
\end{table}

\section{A brief comment on the line shape theory and the challenges with calculating the spectral Far wing} \label{sec:far-wing-challenges} 

The dipole moment of an atom or molecule describes the separation of charge within it. 
The line shape can be formally defined as the Laplace transform\footnote{The Laplace transform is a mathematical technique used to convert a function from the time domain to the frequency domain. By taking the Laplace transform of the dipole autocorrelation function, the line shape can be characterised as a function of frequency} of the thermally-averaged dipole autocorrelation function, which is a measure of the loss of coherence of the radiation due to the environment (perturbing electrons, ions, molecules, …).  The relationship between frequency in the line shape and time in the autocorrelation function shows that the time dependence of certain phenomena will impact different parts of the line shape.   Processes that happen within a certain time $\Delta t$ will appear in the spectrum within a frequency range of  $\Delta f=1/ \Delta t$.  Because small  $\Delta f$ samples large  $\Delta t$, the cores of lines are in the dynamic limit because the perturbing particles move appreciably within a large  $\Delta t$.  In the wings of the lines, large $\Delta f$ corresponds to short $\Delta t$.  At short $\Delta t$ the perturbing particles do not move appreciably and appear ``static''.  Thus, the line wings can be determined by `quasi-static' methods.  For example, \cite{Ma1990Water-millimeter-JCP} developed a theory for the calculation of the continuous absorption due to the far wing contributions of allowed lines of H$_2$O- H$_2$O absorption based on a generalization of Fano's theory  and later extended it using the quasistatic approximation for the far wing limit and the binary collision approximation \citep{ma1991far}.  Their development was limited to binary collisions (first-order in density). \citet{Hussey1975Kinetic} developed expressions containing a generalized Fano operator that has terms containing the effects of three- or four-body collisions for spectral line broadening in plasmas.  These terms would be necessary for high-pressure applications.  The wings of spectral lines can be either sub-Lorentzian or super-Lorentzian. It  is the interaction between the radiating and perturbing molecules in the quasistatic limit that determines the sub- or super-Lorentzian behaviour.
    
The simulation of molecular spectra has become routine and automated since the development of the spectroscopic databases in the early 1970s \citep{AFGL1973}: HITRAN \citep{HITRAN2020}, GEISA \citep{delahaye20212020}, ATMOS \citep{gunson1996atmospheric}, etc.  Here we will use HITRAN as the standard.  However, many problems remain, and therefore these databases and their functionality are periodically updated and extended.  First, there is always the question of the quality of the molecular parameters in these databases.  Positions and lower state energies (term values as the units are cm$^{-1}$) are well-known for many molecules, but the line intensities are relatively known for a number of molecules in mostly low temperature ranges.  The line shape parameters, $\gamma$, and $\delta$ (i.e., the Lorentzian pressure-broadening and pressure-shift parameters, respectively) are the least well-known.  For limited number of the molecules in HITRAN, there are measured half-widths and a small number of shifts.  However, for many cases, these pressure-broadening coefficients are limited to few transitions, and hence they are estimated for all other lines.
The uncertainty in the estimated half-width ranges from $\sim$1-100$\%$, and in some cases, where the experimental or theoretical data is lacking, the values in HITRAN \citep{HITRAN2020} are estimated using different methods. For instance, they can be estimated by scaling data from other collision systems \citep{Gamache2009}, using the mass dependence of half-widths \citep{LAMOUROUX2010}, or using the ratios based on the few available empirical values available in the literature \citep{Tan2022}, or by averaging the values as a function of quantum numbers (for instance, for unmeasured transitions of water vapor in HITEMP2010 \citep{Rothman2010}, or simply a constant value for all transitions.  The line shifts are even less certain as they exhibit much stronger vibrational dependence.

One of the main problems associated with using the Voigt or Lorentz profiles is that the model does not give the correct profile over the whole extent of a spectral line. Burch {\it et al.} \citep{burch1963total, burch1971absorption, burch1968absorption, burch1967absorption, burch1956infrared, burch1962total}  made several studies in the sixties and seventies showing that the Lorentzian line shape does not match measured spectra.  They note for \ce{CO2} ``Beyond a few cm$^{-1}$ from line centre all lines absorb less that Lorentz-shaped lines with the same $\gamma$.''  For water vapor, the opposite is true, the absorption is greater than that modeled by Lorentzian as one moves away from the line centre.  At around the same time, the radiative transfer model FASCODE \citep{smith1978fascode, clough1981atmospheric} was being developed (now it is the LBLRTM \citep{clough2005atmospheric} code). To deal with this problem and to reduce the computation time, line wings were cut off at $\pm$25~cm$^{-1}$ from the line centre.  For broad (low J) lines, of \ce{H2O} and \ce{CO2} at room temperature and 1 atmosphere, this amounts to going out to $\sim$333 half-widths, but note as Burch {\it et al.} have suggested there are noticeable differences in the absorption at $\sim$25 half-widths from the line centre. Note also that for narrow lines (generally high J), going out $\pm$25~cm$^{-1}$ corresponds to many more half-widths.  To correct for this cut-off additional absorption is accounted for by what has been called the continuum, designated as the correction factor or $\chi$ factor. Such a correction factor was used for simulating CO$_2$ emission to compare to observations of the night side of Venus~\citep{93PoDaGr}. 
The continuum is different for each absorbing molecule.  For water vapour, it was based on the work of Clough {\it et al.} and was called CKD \citep{clough1989line, ma1992far} This model has been updated many times based on new measurements and simulations. The current version is called MT$\_$CKD \citep{mlawer2012development}.  The use of the continuum recognizes that the absorption arises from several mechanisms and that corrections need to be made by cutting off the lines at 25 cm$^{-1}$.  First, there is the far-wing collision-induced absorption as modeled by Ma and Tipping \citep{ma1990atmospheric, ma1990water, ma1991far, ma1994detailed, ma1992far, ma1999theoretical}, and there is also absorption by metastable species \citep{22SiPtEl} (dimers, trimers, etc. although most of this can be attributed to the dimers).  For example, the \ce{H2O} absorption can be broken into collision-induced absorption, dimer spectrum, and a correction for the line cut-off.  It will be similar to other molecules.

\section{Challenges}\label{sec:challenges}

\subsection{Computing opacities for high pressures: Line mixing}\label{sec:high-pressure-challenge}


 Spectroscopic studies of exoplanets largely probe regions whose pressure are less
than a few bars and, in many cases, a lot lower than this. For these regions, the Voigt profile is
known to give a reasonable approximation to the pressure and temperature-dependent line shape
in the region of the line centre although subtle collision effects are known to lead to minor
distortions of this \citep{jt584}. At high pressures, there are a number of complications
in simulating the spectrum. Table \ref{tab:comp} lists the T and P grid points for the MAESTRO database. The line shape parameters are determined based on the number density of the perturbing
molecules with the pressure of 1 atm. At high pressures, non-ideal behavior of the gases occurs
and the line shape parameters should be taken at the corrected pressure corresponding to the
number density of the real gas. The correction can be found in the appendix of \citet{Sung2019_O2} who discuss O$_2$; information on compressibility for key molecules can by found in the on-line NIST Chemistry Webbook \footnote{\url{https://webbook.nist.gov/chemistry/}}. However, the largest distortion away from a Voigt profile is caused by line mixing.
Physically line mixing is occurs due to collisions that connect the two transitions  \citep{1992Levy,2008Hartmann, PIERONI2001}.  To see this, consider two transitions that mix, \textit{f} $\leftarrow$ \textit{i} and \textit{f$^{\prime}$} $\leftarrow$ \textit{i$^{\prime}$}.  Line mixing provides an alternative path to go from $i$ to $f$.  A molecule in state \textit{i} undergoes a collisional transition to \textit{i$^{\prime}$}, which then absorbs a photon and goes to \textit{f$^{\prime}$}; \textit{f$^{\prime}$} then undergoes a collisional transition to \textit{f}, thus completing the path from \textit{i} to \textit{f}.  This path amounts to a redistribution of population, which does not affect the line position, but the line intensity and line shape parameters can be greatly affected.  To be efficient, the frequencies of the two transitions should be very close to each other and the probability of the collisional connections, which are given by reduced matrix elements \citep{LAMOUROUX2014, GAMACHE2019},  should be large.  As pressure
increases collision induced absorption (CIA), which depends on the square
of the pressure, becomes increasingly dominant.

Firstly, Sung et al. (2019) discussed it only for one single molecule, O2. One good source of compressibility would be the NIST Chembook on-line. Secondly, what is underlined is not quite true. Collision-Induced Absorption (cia) is density-square dependent while line mixing effect is proportional to density. Therefore, toward high pressures, the cia effect is expected to be significant (~10 bars), substantial (~100 bars), and eventually dominant (~1000 bars) over the ACS based on the Voigt profile.
 The effect
of line mixing with increasing pressure is gradually blending discrete lines into a single, strongly broadened absorption feature (see Fig. IV.2 of \citet{2008Hartmann} for a nice
illustration of this process).
At terrestrial temperatures and pressures, line mixing is only significant in a few special cases where lines lie particularly close together such as Q-branch transitions of CO$_2$. Thus, for example,
the current HITRAN2020 release \citep{HITRAN2020} implements  full and first-order line-mixing for air- and self-broadened CO$_2$ lines \citep{Hashemi_HITRANCO2} and first-order line mixing of N$_2$O and CO lines \citep{Hashemi_HITRANN2OCO}; see also
a recent experimental analysis of line mixing models for CO$_2$ at high temperature and pressure \citep{Cole2023}.

However, models of gas giant exoplanets need to consider radiative transport effects in regions where the pressure is much higher than 1 bar; the {\tt MAESTRO} database provides cross sections
extending up to 3000 bar. In this high-pressure regime, the ideal gas law is not longer obeyed  and line mixing not only becomes important but increasingly becomes the dominant pressure-broadening mechanism. The standard
procedure for modeling line mixing is via the use of  
a frequency-dependent relaxation matrix \citep{1992Levy,2008Hartmann}. Empirical tests of the methodology using the spectrum of NO molecule at pressures of 30 - 100 bar and temperatures
over 500 K \citep{1996Fu,2021Almodovar} suggest that this methodology is reasonably successful at modeling the effects of line mixing. We note that in this pressure -- temperature regime,
the individual transitions merge into single quasi-continuous features more characteristic of a condensed phase than the usually heavily-structured line spectra which characterize gas phase spectra. In this context, we note that in the high-pressure regime of interest for the interior of gas giant (exo)-planets the species generally exist in an environment well above their triple point meaning that there is no longer a distinction between gas and liquid phases. It is therefore not altogether surprising that the resulting spectra take on some of the characteristics of a condensed phase spectrum.

Very recently, \citet{Ren2023} performed an experimental study of high-pressure absorption coefficients for CO$_2$, CO, and H$_2$O for pressures up to 80 bar.  They found that an empirical pseudo-Lorentz line shape model, which considers line-mixing effects is easy to implement and relatively accurate at modeling high-pressure gas spectra for the conditions they tested. This suggestion certainly merits further investigation.

We are not aware of any exoplanetary models which allow for the effects of line mixing. This is not surprising as the explicit coupling of the very many transitions that occur in a hot spectrum
via a relaxation matrix which links these transitions together represents a significant extra complication in the radiative transport problem. It is likely that this problem is best
addressed by generating effective temperature and pressure dependent cross section sets which include line-mixing effects which can simply be used in the appropriate models.
This idea is left to future work. The onset of high-precision JWST spectra as well as high-precision medium-resolution ground-based spectra of Brown Dwarfs and giant planets will enable constraints on pressure-temperature profiles to pressures of $\sim$1-10 bar \cite[e.g.][and as seen in Figure \ref{fig:TPprofile-ContributionFunc}]{Hood2023}. Therefore, we will soon be able to shed light on whether line mixing seems to be driving temperature structure by comparing radiative-convective-derived models to retrieved temperature profiles.


\subsection{Spectral Windows regions}\label{sec:spectral-windows}

The question of how to limit the extent of the line wings as a function of pressure and temperature is complex. At higher pressures, in order to capture the majority of the opacity in any given line transition it may be necessary to extend the line wings far beyond the line core and thus far beyond the validity of the Lorentzian profile. The lack of an extension will result in underestimating the total opacity, which will predominantly effect regions of low opacity --- so-called ``spectral windows''. Because these regions of low opacity  contribute significantly to the character of the emerging radiation, they have to be modeled correctly. The areas of the spectrum that are already
optically thick will not be affected to any great degree by decisions to modify the line wing extent. However because the far wings in many cases are known to be very sub-Lorentzian, the correct approach to the modeling of the spectra is complicated. One approach is to establish a cutoff [in cm$^{-1}$]
from the line centre for the profile and then attempt to compensate for the missing line wings by a pseudo-continuum contribution that would have to be computed separately. Although this approach is already used in many applications related to the Earth's atmosphere for water opacity (as discussed in \S\ref{sec:far-wing-challenges}), how this would be handled in the astronomical case is less clear. A recent study by \cite{22AnChEl} found that transmission spectra of exoplanet atmospheres which are close to Earth's atmospheric temperature can be significantly impacted if data for the water continuum is not included in models. The significance of the continuum is less clear at higher temperatures, but is generally assumed to follow a negative temperature dependence~\citep{16ShCaDi}. Exoplanetary and brown dwarf atmospheric radiative-transfer models cover a wide range of temperatures and
pressures that are never encountered on the Earth and include many species whose
spectra may not be fully documented.

\section{Case studies}

 \subsection{Benchmark the Effect of Different wing cut-off Methods on \ce{CH4}-in-air System}\label{sec:Benchmark-CH4-cutoff}
 
\begin{figure}[bh!]

	\includegraphics[scale=.8]{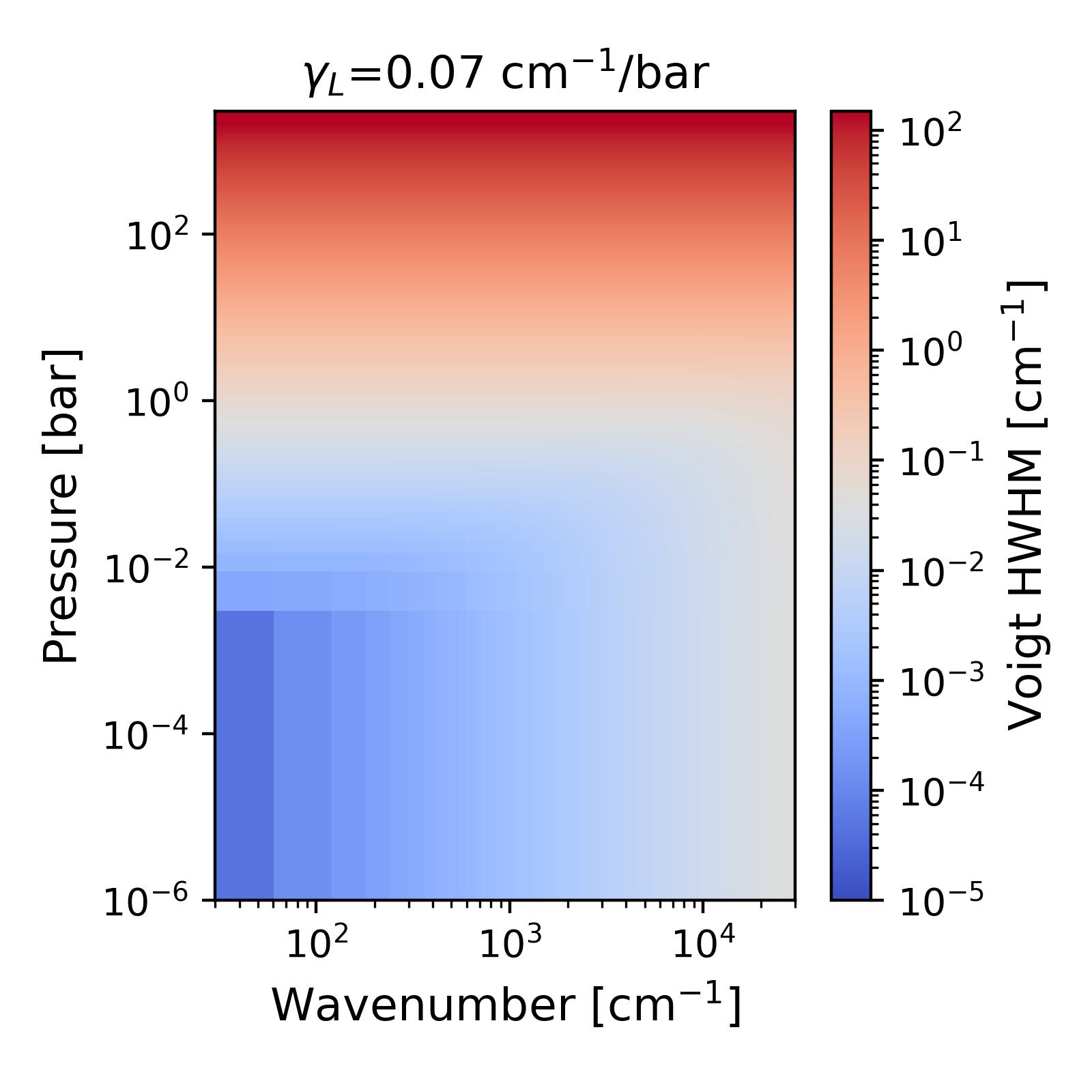}
    \caption{ Estimated Voigt HWHM across a wide range of spectral range and pressure for \ce{CH4} broadened by air with a constant Lorentz coefficient of 0.07 cm$^{-1}$/bar and for room temperature. The calculated Voigt HWHM varies smoothly with pressure and position, the apparent steps are an artifact of the number of points calculated. 
    }
    \label{fig:Voigt_HWHM_WN_P}
\end{figure}

Given the challenges on the theory side as well as the lack of sufficient experimental spectroscopic parameters, our goal is to propose the standard practice for calculating the wing cut-off for the astronomy community. But first, we need a quantitative overview of the range of Voigt HWHM (i.e., $\Gamma_{\rm V}$) values. Fig.~\ref{fig:Voigt_HWHM_WN_P} provides this information across a wide range of pressures and wavenumbers for room temperature
\footnote{Note that the Voigt HWHM ($\Gamma_V$) in Fig.~\ref{fig:Voigt_HWHM_WN_P} is calculated with the following equation:  $\Gamma_{\rm V} = 0.5346 \enspace \Gamma_{\rm L} +  \sqrt{0.2166  \enspace \Gamma_{\rm L}^2 + \Gamma_{\rm G}^2} $ where $\Gamma_{\rm G}$ is the Doppler HWHM in cm$^{-1}$   \citep{Olivero1977}. }. 
For the sake of simplicity in comparison,  a constant Lorentz coefficient of $\gamma_L$~=~0.07~cm$^{-1}$/bar is used here but in reality, $\gamma_L$ is dependent on the molecular symmetry and quantum numbers of the transition. In this figure shows that $\Gamma_{\rm V}$ HWHM has a wide range of $10^{-1}$--150~cm$^{-1}$ across the pressure-wavenumber space. Cases with very low pressures ($\lesssim10^{-2}$ bar) and low wavenumbers $\lesssim10^{3}$ cm$^{-1}$ (or $\gtrsim$~10 \micron), have $\Gamma_{\rm V}$ lower than 10$^{-2}$ cm$^{-1}$. On the other hand, very high pressures ($\gtrsim$10$^{1}$ bar) lead to a large  $\Gamma_{\rm V}$ HWHM, i.e., ~>~20~cm$^{-1}$, and up to $\sim$150~cm$^{-1}$.

\begin{figure*}

	\includegraphics[scale=.77]{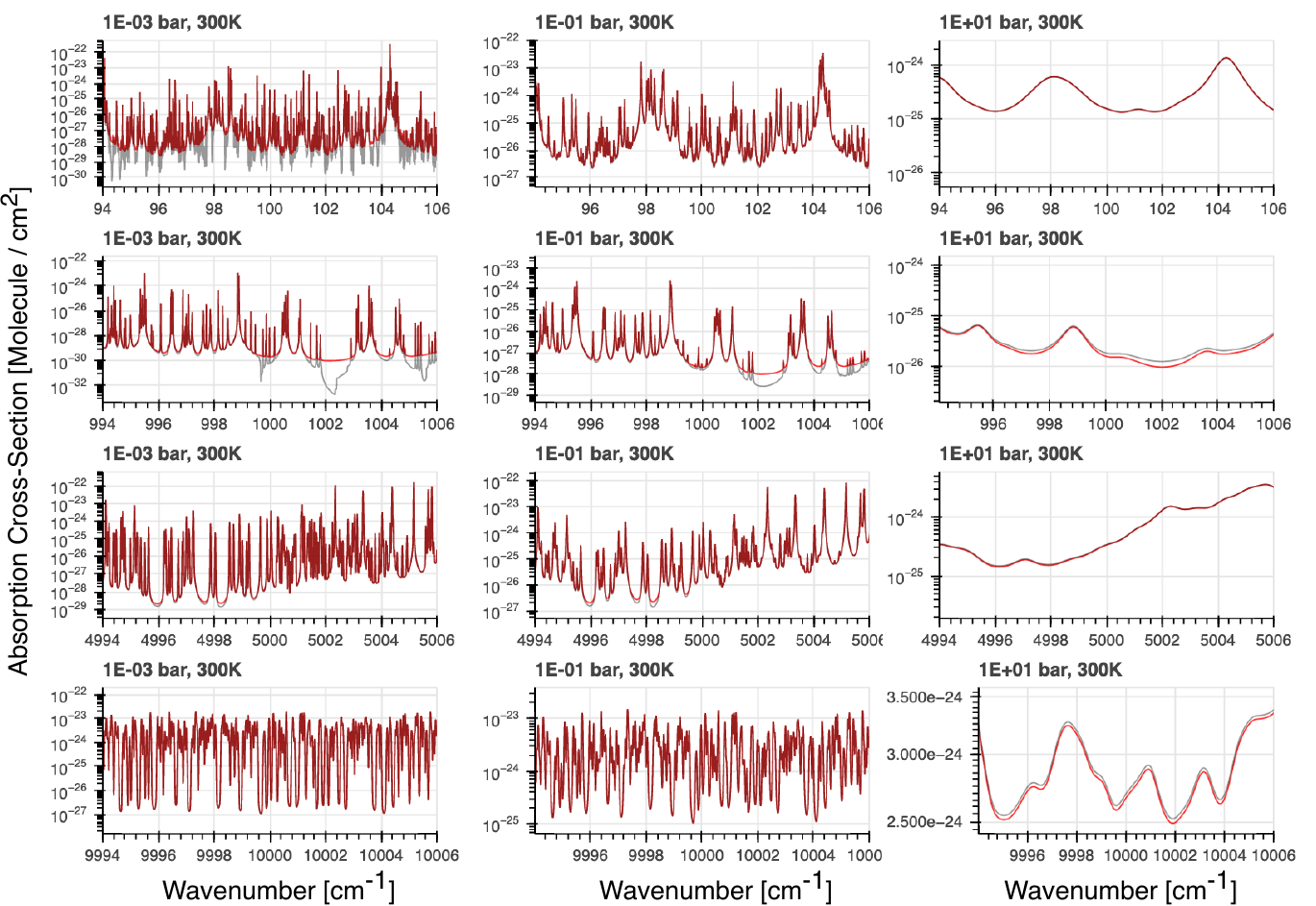}
    \caption{ Calculated cross sections of \ce{CH4} at 300~K broadened by air at three sample pressures (0.001, 0.1, 10.0 bar) in four spectral regions, using fixed line wings ($R_{\rm cut,Abs}$, (red spectra) and line wings dependant on the pressure-broadening ($R_{\rm cut,HWHM}$, grey spectra) represented in Eqs.~\ref{eq:R_cut_senario-Abs}-\ref{eq:R_cut_senario-HWHM}.}
    \label{fig:ACS_all_spectra_hitran_ch4_air}
    
\end{figure*}

\begin{figure*}

	\includegraphics[scale=.6]{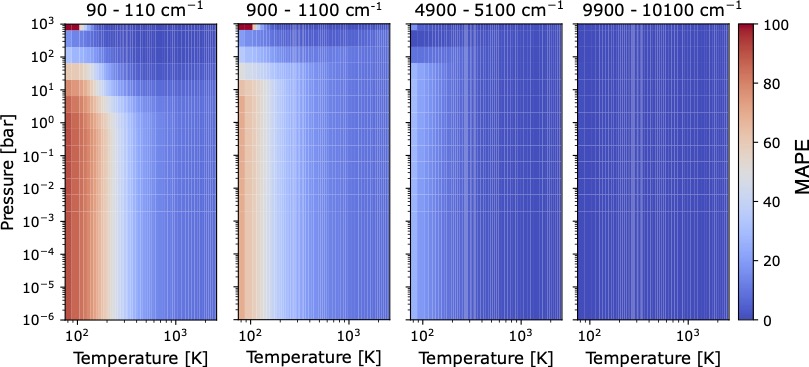}
    \caption{ Mean Absolute Percentage Error (MAPE) from the two wing cut-off approaches for different spectral ranges: From left to right: $\sim$100 \micron, $\sim$10 \micron, $\sim$2 \micron, and $\sim$1 \micron for \ce{CH4} in air. This figure shows that   the generated opacities at low wavenumbers, low pressures, and low temperatures are highly sensitive to the spectral wings and wing cut-off approaches. The lack of broadening coefficients and appropriate line profile for very high pressures is another issue.  See section \S\ref{sec:Benchmark-CH4-cutoff} for further details.
    }\label{fig:compare_Rcutoff_senarios_MAPE}
\end{figure*}

Fig.~\ref{fig:ACS_all_spectra_hitran_ch4_air} represents the generated ACS datasets for the \ce{CH4}-in-air system for a wide range of pressures and temperatures, 10$^{-6}$~--~10$^3$~bar and 75~--~4000~K, and for four wavenumber (WN) regions: 90~--~110, 900~--~1100, 4900~--~5100, and 9900~--~10100~cm$^{-1}$ (or 100, 10, and 2, 1 \micron, respectively) with $R_{\rm cut}$ calculated using in Eqs~(\ref{eq:R_cut_senario-Abs}) and (\ref{eq:R_cut_senario-HWHM}). To quantify the impact of the wing cut-off, Fig.~\ref{fig:compare_Rcutoff_senarios_MAPE} represents the differences in MAPE (the mean absolute percentage error) metric between the ACS datasets for hundreds of spectra.

\begin{equation}\label{eq:R_cut_senario-Abs}
    R_{\rm cut,Abs} = 
\left\{\begin{matrix}
 30~\textrm{cm}^{-1}~\ \ \textrm{for P}~\leqslant 200~\textrm{bar} \\ 
 150~\textrm{cm}^{-1}~\textrm{for P} > 200~\textrm{bar} 
\end{matrix}\right.
\end{equation}

\begin{equation}\label{eq:R_cut_senario-HWHM}
    R_{\rm cut,HWHM} = 500\times\gamma_{\rm Voigt}   
\end{equation}

If we take the $R_{\rm cut, Abs}$ as the actual value, then  MAPE formula is expressed as follows: 
\begin{equation}\label{eq:mape}
        \textrm{MAPE} = \left(\frac{100}{n_{\rm gp}}\right)\sum_{i=1}^{\nu_{\rm gp}} \left| \frac{ \sigma_{R_{\rm cut,Abs}(\nu,T,P)} - \sigma_{R_{\rm cut, HWHM}(\nu,T,P)}}{\sigma_{R_{\rm cut,Abs}(\nu,T,P)}} \right| 
\end{equation}
where $n_{\rm gp}$ is the number of grid points in the pressure-temperature space and $\nu_{\rm gp}$ is the wavelength of the transition. Few examples of the generated ACS spectra from the two scenarios of the wing cut-off are represented in Fig.~\ref{fig:ACS_all_spectra_hitran_ch4_air} for T= 300~K and pressures 10$^{-3}$, 10$^{-1}$, 10~bar and wavelength 90~--~110, 490~--~510, 990~--~1010, and 9990~--~10010~cm$^{-1}$. We see the largest discrepancies in the generated ACS are from cases with low pressures, P < 0.1, and low wavenumbers.


Following that, Fig.~\ref{fig:compare_Rcutoff_senarios_MAPE} illustrates the differences between these two scenarios in the pressure--temperature space and across a wide spectral range. The balance between the Doppler widths and Lorentz widths plays a key role in the MAPE amount between these two $R_{\rm cut}$ scenarios. This difference is $>$50$\%$ for very low wavenumbers (i.e., WN$\leq$1000cm$^{-1}$ or at far-infrared $\lambda$>10\micron). MAPE error becomes less than 10$\%$ for high wavenumbers. The lesson learned from this figure is that the generated opacities at low wavenumbers, low pressures, and low temperatures are highly sensitive to the spectral wings and wing cut-off approaches. This is in agreement with the findings of \citet{Niraula2023} where the effect (or lack of effect) of pressure broadening and far-wing on the retrievals from the WASP 39b spectra (which probed the low-pressure/high-temperature layer of the atmosphere) was investigated. Please note that this figure serves as an illustration of the magnitude of the MAPE error for this specific absorber. For absorbers with different molecular masses (and consequently different Doppler widths), the errors would differ accordingly.  

The current focus of exoplanet and brown dwarf atmospheric spectroscopy with JWST is on the spectral range at wavenumbers $>500$~cm$^{-1}$. For example, the very first JWST Early Release Science observations focused on exoplanets and brown dwarfs featured a 1-20~$\micron$ R$\sim$1000-3700 spectrum of VHS 1256-1257 b \citep{Miles2022}, and a 1-5$\micron$ R$<300$ spectrum of WASP-39 b \citep{jwst2023identification}. Additionally, radiative-convective climate modeling demands accurate opacities throughout the expected spectral energy distribution of the planet's emitted flux. Figure 3 in \citet{mukherjee2023} shows typical wavelength bins used for radiative-convective models of planets with T$_\mathrm{eff}=300-1000$ K. These wavelength bins extend to 30~cm$^{-1}$ (333~$\micron$). Therefore, from an observational and heat transport perspective, it is essential to establish uniform methods of computing wing cut-offs.

\subsection{Benchmark the Effect of Different wing cut-off Methods on CO$_2$-in-\ce{N2} System}\label{sec:lab_spectra_co2}

The strong $\nu_{3}$ vibrational mode of CO$_{2}$ near 2350\,cm$^{-1}$ (4.25~$\mu$m) provides an opportunity to test the impact of the line wing cut-off over a complete absorption band. This band is readily observed in terrestrial atmospheres and has recently been detected in JWST observations of WASP-39b \citep{jwst2023identification}. 

\begin{figure*}
  \includegraphics[trim={0 0 0 0cm},clip,width=0.97\linewidth]{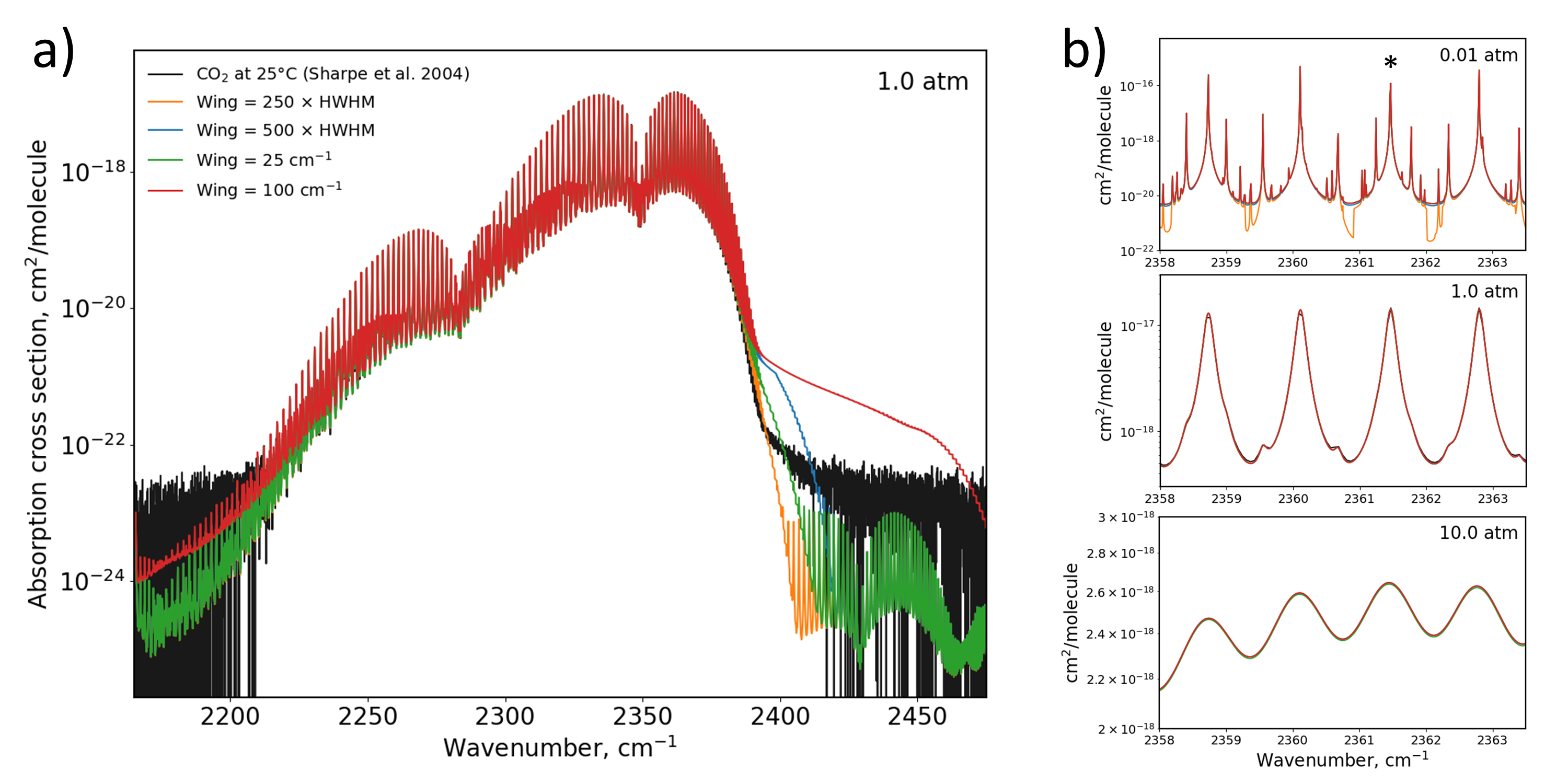}
\caption{A demonstration of the effect of the line wing cut-off for the $\nu_{3}$ region of CO$_{2}$. a) An experimental absorption cross-section of CO$_{2}$ at 25$^{\circ}$C broadened by 1.0\,atm of N$_{2}$ recorded by the Pacific Northwest National Laboratory, PNNL \citep{PNNL-REF}, compared to four calculated spectra with different line wing cut-off limits. b) A zoomed in comparison at 25$^{\circ}$C for three pressures. Note, only the plots at 1.0 atm include the PNNL spectrum. See Section~\ref{sec:lab_spectra_co2} for further details. }
\label{fig:CO2vPNNL}
\end{figure*}

Figure~\ref{fig:CO2vPNNL} compares the effect of the line wing cut-off for the $\nu_{3}$ region of CO$_{2}$. The left panel displays an experimental ACS of CO$_{2}$ at 25$^{\circ}$C broadened by 1.0\,atm of N$_{2}$ recorded by the Pacific Northwest National Laboratory, PNNL \citep{PNNL-REF}. The experimental measurement (black line) is shown in the region of the strong $\nu_{3}$ vibrational mode near 2350\,cm$^{-1}$ on a logarithmic scale. The approximate sensitivity of the experimental measurement is $\sim$10$^{-23}$~cm$^{2}$/molecule. Overplotted are four spectra calculated with Voigt line shapes using HAPI \citep{HAPI-REF} and the CO$_{2}$ line list from HITRAN2020 \citep{HITRAN2020}. The only difference between the four calculated spectra is the treatment of the line wing cut-offs. These include line wings calculated from each line centre in multiples of the Voigt HWHM (250$\times$ and 500$\times$) as well as fixed widths (25 or 100\,cm$^{-1}$). The effect of the different line wing treatments is clearly seen in the region >~2380~cm$^{-1}$. 

The right panel of Figure~\ref{fig:CO2vPNNL} shows a zoomed-in region near the centre of the band for three different pressures (the example at 1.0~atm is the same as the left panel). At lower pressures, the limitation of calculating a line wing that is dependent on the line broadening is shown to produce unexpected steps in the ACS between the strongest lines. In this example, these steps are a consequence of calculating the line wings up to 250$\times$HWHM. This can be seen for the line indicated by an asterisk, where at 0.01~atm the broadening is 0.002~cm$^{-1}$, which results in line wings of approximately $\sim$0.5~cm$^{-1}$ either side of the line centre.

It should be noted that the comparisons in Figure~\ref{fig:CO2vPNNL} use the air-broadening half-widths from HITRAN. For a more accurate comparison to the PNNL spectra, it would be necessary to use N$_{2}$-broadening parameters along with line-mixing. However, the effects observed by the different choices in line wing cut-off would still be observed.

\subsection{Assess the Impact of the Super-Lines Method on wing cut-off: \ce{SO2}-in-\ce{H2}/He system} \label{sec:superlines}

Here we assess the effect of using the super-lines method to compute absorption cross-sections as a comparison to the effect of line-wing cut-off.
Some line lists are very large and can contain billions of transitions between many millions of energy states. When computing cross-sections from such large line lists, studies such as \cite{TheoReTS} and \cite{jt698} have employed so-called super-lines. 
The general principle 
is to first compute line intensities $I_{ij}$ at a given temperature (i.e. stick spectra) and then to sum together the intensities of all lines within a specified spectral bin to create a so-called super-line. Thus super-lines are effectively temperature-dependent spectral histograms computed on a spectral grid. In order to ensure error transmission is kept to a minimum,  a non-uniform grid of R~=~1,000,000 is typically used for the first step of the super-lines computations with a typical size of a few million grid points (super-lines) for each of the 20--30 temperatures used.
A Voigt broadening profile is then applied to each of these super-lines. Since the computation of the Voigt  profile is the most expensive part of a simulation, this two-step procedure provides a huge reduction in the computation time. 
 
The main source of error of the super-line procedure is that any dependence of the pressure broadening line profiles on the quantum numbers and molecular symmetry is ignored. This is because each super-line is a combination of multiple transitions connecting different states  falling into the given spectroscopic bin.

 Here we consider the high-resolution cross-sections computed for ExoMolOP~\citep{20ChRoAl}, prior to sampling down to sampled cross-sections or k-tables\footnote{The general principle of k-tables is to order spectral lines within a given spectral bin, producing a smooth cumulative distribution function to represent opacity. This smooth distribution can be more efficiently sampled, allowing k-tables to reach comparable accuracy as sampled cross-sections for a much smaller R~=~$\frac{\lambda}{\delta\lambda}$.}.
 We compare the cross-sections of SO$_2$, using the ExoCross~\citep{ExoCross} code with the ExoMol ExoAmes line list of \cite{jt635}, as an illustrative example, with and without using the super-lines method. The super-lines methods necessitate that average-$J$ rather than $J$-dependent broadening parameters are used. 
 The SO$_2$ line list of \cite{jt635} contains 1.4~billion transitions between 3.3~million ro-vibrational energy levels and covers up to 8000~cm$^{-1}$ (down to 1.25~$\mu$m).
 We use average broadening parameters (i.e. not $J$-dependent) for the super-lines method, compared to $J$-dependent broadening parameters extracted from the HITRAN database~\citep{HITRAN2020}
 for the broadening of SO$_2$ by H$_2$ and He. 
 A fixed line wing cut-off of 25~cm$^{-1}$ was used for both the super-lines and the standard non-super-lines methods.
 A variable grid spacing is used to try and ensure at least 4 sampling points for each line, with a maximum separation of 0.01~cm$^{-1}$.

\begin{figure*}
	\includegraphics[scale=0.7]{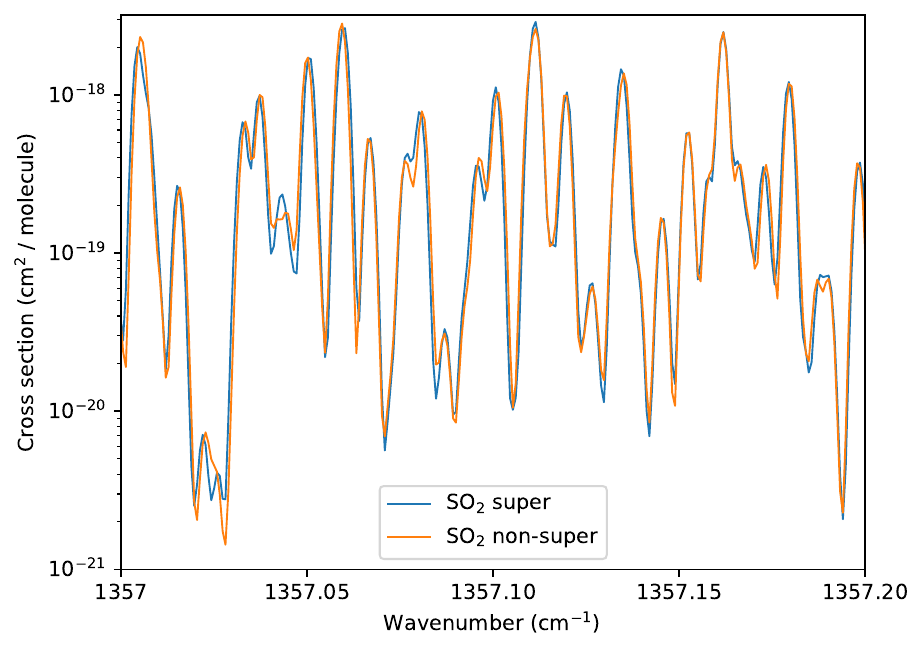}
    \caption{ An example of cross-sections of SO$_2$ computed using ExoCross~\citep{ExoCross}, with the super-lines method compared to without, at 500~K and 1~$\times$~10$^{-5}$~bar. See section \S\ref{sec:superlines} for further details.
    }
    \label{fig:super_comp_lowPT}
\end{figure*}

 To compute the mean absolute percentage error (MAPE), we use the metric of Eq.~(\ref{eq:mape}). Here we take the non-super-lines approach as the actual value, where we also use a fixed line wing cut-off of 25~cm$^{-1}$. We focus on the region 1300~-~1400~cm$^{-1}$ (7.1~-~7.7~$\mu$m). The MAPE value between the non-super-lines and the super-lines method for a selection of pressures and temperatures using SO$_2$ as an example is summarized in Table~\ref{t:mape_super}. 
 It can be seen that the error increases with lower pressures and temperatures, where the lines are narrower than at higher pressures and temperatures. An example of cross-sections of SO$_2$ computed using the two methods at 500~K and 1~$\times$~10$^{-5}$~bar can be seen in Figure~\ref{fig:super_comp_lowPT}.

 	\begin{table}
 	\caption{Mean absolute percentage error (MAPE) of the super-lines vs non-super-lines method for computing cross-sections of SO$_2$ for a selection of pressures and temperatures. }
 	\label{t:mape_super} 
 	\centering  
 	\begin{tabular}{lcc}
 		\hline
 		\hline
 		\rule{0pt}{3ex}Temperature (K)  & Pressure (bar) & MAPE ($\%$)\\
 		\hline
 		\rule{0pt}{3ex} 1500 &  1 &  7.8$\times$10$^{-4}$\\
 		\rule{0pt}{3ex} 1500 &  1$\times$10$^{-2}$ &   0.09\\
 		\rule{0pt}{3ex} 1500 &  1$\times$10$^{-5}$  & 0.18 \\
 		 		\rule{0pt}{3ex} 1000 &  1 & 5.90$\times$10$^{-3}$\\
 		\rule{0pt}{3ex} 1000 &  1$\times$10$^{-2}$ &0.21 \\
 		\rule{0pt}{3ex} 1000 &  1$\times$10$^{-5}$  & 0.83 \\
 		 		\rule{0pt}{3ex} 500 &  1 & 0.03\\
 		\rule{0pt}{3ex} 500 &  1$\times$10$^{-2}$ & 0.03\\
 		\rule{0pt}{3ex} 500 &  1$\times$10$^{-5}$  & 6.34\\
 		\hline
 		\hline
 	\end{tabular}
\end{table}

 The cross-sections computed using the super-lines technique (which necessitates the use of average-$J$ broadening parameters) have negligible differences 
 when compared to those computed without, while they have a huge saving in efficiency. This is particularly true for large line lists with many millions of levels and billions of transitions. The differences appear to be much smaller than the typical differences found between using different wing cut-off methods. The latter is illustrated by Figure~\ref{fig:compare_Rcutoff_senarios_MAPE}. 
 We assess that cross-sections computed with the use of super-lines are appropriate for low- to mid-resolution applications, such as modeling JWST observations. As long as each of the super-lines is adequately sampled then the overall opacity will largely be conserved. 

\section{Our recommendation for choosing the Wing cut-off}\label{sec:recommendation}

The scope of our paper does not encompass resolving current challenges such as line mixing, non-Lorentzian behavior, or inaccuracies in various Voigt algorithms. Our primary objective is to establish a standardized approach for opacity calculations within the exoplanetary and astrophysical communities, specifically focusing on wing cutoffs.
For the systems where the empirical parameterizations of far-wing exist, they should be used, although one has to be aware that there is still room for improvement. 
Further theoretical studies and spectroscopic measurements are still required to create comprehensive insights into the physics of the far wings across the pressure regime and within the different mixtures of atmospheric gases that haven't been studied yet. However, given the different scenarios, we examined in Figs.~\ref{fig:ACS_all_spectra_hitran_ch4_air}--\ref{fig:CO2vPNNL}, we propose the following method, assuming a Voigt profile, to calculate the wing cut-off for these pressure ranges: 

\begin{equation}\label{eq:R_cut_best_practise}
    R_{\rm cut,Abs} = 
\left\{\begin{matrix}
 25~\textrm{cm}^{-1}~\ \ \textrm{for P}~\leqslant 200~\textrm{bar} \\ 
 100~\textrm{cm}^{-1}~\textrm{for P} > 200~\textrm{bar} 
\end{matrix}\right.
\end{equation}

This method is chosen for the following reasons :

\begin{enumerate}
    \item For intermediate pressures and high wavenumber (>~5000~cm$^{-1}$), the two common methods in Eqs.~\ref{eq:R_cut_senario-Abs} and \ref{eq:R_cut_senario-HWHM} are in a very good agreement. However, for very low pressures (<~10$^{-1}$) and for some spectral ranges, they result in some disagreement as shown in Figs.~\ref{fig:ACS_all_spectra_hitran_ch4_air}--\ref{fig:CO2vPNNL}. At low pressures, the Voigt HWHM is $<0.01$cm$^{-1}$ (see Fig.~\ref{fig:Voigt_HWHM_WN_P}) and the values calculated using \ref{eq:R_cut_senario-HWHM} are physically small, therefore using fixed values is beneficial. 

    \item For very high pressure i.e., >~200 bars, the Voigt HWHM is more than 20~cm$^{-1}$ and so to model the core of the line profile and a part of the wing, an absolute value of 100~cm$^{-1}$ is proposed. It is worthwhile to note that this value might seem very small for P~>~1000 bar, but the lack of sufficient thermodynamics and spectroscopic studies in this high-pressure regime forced us to conservatively choose this number of 100~cm$^{-1}$ (see \S\ref{sec:high-pressure-challenge}). 

    \item A line wing of 25\,cm$^{-1}$ is typically used for calculations of terrestrial radiative transfer \citep{mlawer2012development}. Hence, our proposed method is very well aligned with the planetary community and could be utilized for their modeling studies as well.
\end{enumerate}

\section{Instruction on how to control $R_{ cut}$} \label{sec:codes}
The pre-generated ACS data in the current version of {\tt MAESTRO} were computed using HAPI and ExoCross codes. Hence, you can see a short description on how to control the $R_{\rm cut}$ in these tools. 

\subsection{HAPI}
The HITRAN Application Programming Interface (HAPI) \citep{HAPI-REF} is a set of Python libraries that allow users to download the HITRAN data and carry out different calculations, including absorption cross-sections, emission, transmission, etc. HAPI has a number of line shape functions, and complex probability functions implemented. The users can select among these functions or provide their own. One can also specify the wing cut-off. Although the default value is 50 HWHM, for the applications to planetary atmospheres under pressures encountered in the terrestrial atmosphere, it is recommended to use 25 cm$^{-1}$. Going outside of the terrestrial conditions, one can implement user-defined functions on a case-by-case basis. We will implement recommendations from this paper in future HAPI distributions. In any case, the flexibility of the open-source code gives all power to the user to implement different line shape functions. 

\subsection{ExoCross}

ExoCross \citep{ExoCross} is a Fortran 2003 open-source  program to compute spectra of molecules developed and maintained by the ExoMol project \citep{20TeYuAl}. ExoCross provides several line profiles as well as algorithms to compute them. Apart from the main-stream algorithm  by \citet{79Humlic}, two fast Voigt methods are also provided in ExoCross: an approximated vectorized Voigt and a binned Voigt profile with analytical integrals. Different line profiles can be easily implemented in ExoCross.

The default wing cut-off value is 25~cm$^{-1}$ which however can be changed by the user. 
A python version of ExoCross, PyExoCross has also recently been developed \citep{PyExoCross}
and treats the wing cut-offs similarly.

\section{Conclusions}\label{sec:conclusions}

The main goal of this study is to provide a standard method for controlling the spectral line wing cut-off. Inconsistencies, potential inaccuracies/biases, and the lack of necessary theoretical studies or laboratory measurements are reviewed in this work to highlight the difficulties involved in implementing line shapes and wing cut-off relevant to modeling extrasolar atmospheres.

The generated absorption cross-sections can be very sensitive to the choice of wing cut-off, particularly in certain pressure and temperature ranges across the wavelength space. With the incredible advancements in space exploration over the last decade and a large number of detected exoplanets and brown dwarfs, it is crucial to utilize accurate and standardized cross-sections across different research groups. Our main objective in this study is to promote consistency in atmospheric modeling results by recommending a ``standard'' and ``default'' method for the wing cut-off. The recommendations and discussions provided here aim to provide insights into the applications of the MAESTRO database, but they are also valid and applicable across the fields of astronomy and spectroscopy.

Given the spectral parameters, the simulation of a spectrum has a number of other complications, most are associated with the line shape.  Some molecules exhibit line-coupling effects (or line-mixing effects), which can greatly change the simulated spectra.  There are theoretical models for each mechanism.  However, the main challenge to simulate a spectrum is the choice of a line-shape model.  Most of the line-shape data on HITRAN is for the Lorentz profile; in applications, this is coupled with Doppler broadening to create the Voigt profile. The suggested model is the Hartmann-Tran Profile \citep{tran2013efficient, tran2014erratum,jt584}, but little data are available, especially for large-scale simulations.  There are parameters available for a small number of transitions for other more sophisticated line-shape models.  

The lack of pressure broadening data for pressure and temperature ranges relevant to exo-atmospheres is another issue. Most laboratory spectroscopic measurements are limited to Terrestrial and Jovian atmospheric temperature ranges ($\le$350 K) and up to a few tens of bars. At high pressures, the assumption of the Voigt profile (binary collisions between absorbers and broadeners) is not valid, and more sophisticated line shapes are required to account for non-binary collisions and the intensive interactions that exist between these species.

Therefore, for now, we recommend the default or standard practice, in particular with regards to the {\tt MAESTRO} (Molecules and Atoms in Exoplanet Science: Tools and Resources for Opacities) database, to be the use of a Voigt profile with a line-wing cut-off of 25~cm$^{-1}$ for pressures of 200~bar or less, and 100~cm$^{-1}$ for pressures above 200~bar. Our main reasons for this choice are outlined in Section~\ref{sec:recommendation}. We hope future theoretical and laboratory work will facilitate more precise line profiles for a wide range of species, which can, in turn, be easily implemented in codes to compute cross-sections and thus included in future releases of {\tt MAESTRO}. Having a current ``standard'' or ``default'' recommendation will at least allow for a consistency and thus fair comparison between different studies. Such a comparison is particularly important for the open-access {\tt MAESTRO} database, as it will allow for an easier peer review process\footnote{\href{https://github.com/maestro-opacities/submit-data}{github.com/maestro-opacities/submit-data}} for allowing community contributions of opacity data.

\section*{Acknowledgements}

N.B. \& E.G.N. both acknowledge support from the NASA Astrophysics Division.
E.G.N., N. B.,  \& B. G. acknowledge support from the NASA XRP funding NNH21ZDA001N-XRP. 
N.E.B, N.K.L, I.E.G., R.J.H. et al. {\tt MAESTRO} Grant (NASA Grant Number 80NSSC19K1036).
K.L.C acknowledges funding from STFC under project number ST/V000861/1.
S.N.Y. and J.T. acknowledge supported by the European Research Council (ERC) under the European Union’s Horizon 2020 research and innovation programme through Advance Grant number 883830 (ExoMolHD). We also would like to thank the anonymous reviewers for their insightful comments.

\section*{Software}
HAPI \cite{HAPI-REF}\\
ExoCross \citep{ExoCross} which is available from https://github.com/ExoMol/ExoCross.

\section*{Author Contributions}
All authors have contributed equally to this study.

\section*{Data Availability}
The data underlying this article will be shared on reasonable request to the corresponding author.



\bibliographystyle{rasti}
\bibliography{rasti_template} 

\begin{thebibliography}{88}
\expandafter\ifx\csname natexlab\endcsname\relax\def\natexlab#1{#1}\fi

\bibitem[Almodovar et~al.(2021)Almodovar, Su, Choudhary, Shao, Strand, \&
  Hanson]{2021Almodovar}
Almodovar, C.~A., Su, W.-W., Choudhary, R., Shao, J., Strand, C.~L., \& Hanson,
  R.~K., 2021.
\newblock {Line mixing in the nitric oxide R-branch near 5.2 $\mu$m at high
  pressures and temperatures: Measurements and empirical modeling using energy
  gap fitting}, {\it J. Quant. Spectrosc. Radiat. Transf.\/}, {\bf 276},
  107935.

\bibitem[Anisman et~al.(2022)Anisman, Chubb, Elsey, Al-Refaie, Changeat,
  Yurchenko, Tennyson, \& Tinetti]{22AnChEl}
Anisman, L.~O., Chubb, K.~L., Elsey, J., Al-Refaie, A., Changeat, Q.,
  Yurchenko, S.~N., Tennyson, J., \& Tinetti, G., 2022.
\newblock Cross-sections for heavy atmospheres: H2o continuum, {\it Journal of
  Quantitative Spectroscopy and Radiative Transfer\/}, {\bf 278}, 108013.

\bibitem[Burch \& Gryvnak(1971)]{burch1971absorption}
Burch, D.~E. \& Gryvnak, D.~A., 1971.
\newblock Absorption of infrared radiant energy by {CO$_2$} and {H$_2$O}, {V.
  A}bsorption by {CO$_2$} between 1100 and 1835 cm$^{-1}$ (9.1--5.5 $\mu$m),
  {\it JOSA\/}, {\bf 61}, 499--503.

\bibitem[Burch \& Williams(1962)]{burch1962total}
Burch, D.~E. \& Williams, D., 1962.
\newblock Total absorptance of carbon monoxide and methane in the infrared,
  {\it Applied Optics\/}, {\bf 1}, 587--594.

\bibitem[Burch et~al.(1956)Burch, Howard, \& Williams]{burch1956infrared}
Burch, D.~E., Howard, J.~N., \& Williams, D., 1956.
\newblock Infrared transmission of synthetic atmospheres.* v. absorption laws
  for overlapping bands, {\it JOSA\/}, {\bf 46}, 452--455.

\bibitem[Burch et~al.(1963)Burch, France, \& Williams]{burch1963total}
Burch, D.~E., France, W.~L., \& Williams, D., 1963.
\newblock Total absorptance of water vapor in the near infrared, {\it Applied
  Optics\/}, {\bf 2}, 585--589.

\bibitem[Burch et~al.(1967)Burch, Gryvnak, \& Patty]{burch1967absorption}
Burch, D.~E., Gryvnak, D.~A., \& Patty, R.~R., 1967.
\newblock Absorption of infrared radiation by {CO$_2$} and {H$_2$O}.
  experimental techniques, {\it JOSA\/}, {\bf 57}, 885--895.

\bibitem[Burch et~al.(1968)Burch, Gryvnak, \& Patty]{burch1968absorption}
Burch, D.~E., Gryvnak, D.~A., \& Patty, R.~R., 1968.
\newblock Absorption of infrared radiation by {CO$_2$} and {H$_2$O}. {II.
  A}bsorption by {CO$_2$} between 8000 and 10 000 cm$^{-1}$ (1--1.25 microns),
  {\it JOSA\/}, {\bf 58}, 335--341.

\bibitem[Chubb et~al.(2021)Chubb, Rocchetto, Al-Refaie, Waldmann, Min, Barstow,
  Molli{\'e}re, Phillips, Tennyson, \& Yurchenko]{20ChRoAl}
Chubb, K.~L., Rocchetto, M., Al-Refaie, A.~F., Waldmann, I., Min, M., Barstow,
  J., Molli{\'e}re, P., Phillips, M.~W., Tennyson, J., \& Yurchenko, S.~N.,
  2021.
\newblock The {ExoMolOP Database: C}ross-sections and k-tables for molecules of
  interest in high-temperature exoplanet atmospheres, {\it A\&A\/}, {\bf 646},
  A21.

\bibitem[Clough et~al.(1981)Clough, Kneizys, Rothman, \&
  Gallery]{clough1981atmospheric}
Clough, S.~A., Kneizys, F.~X., Rothman, L.~S., \& Gallery, W.~O., 1981.
\newblock Atmospheric spectral transmittance and radiance: Fascod1 b, in {\em
  Atmospheric transmission\/}, vol. 277, pp. 152--167, SPIE.

\bibitem[Clough et~al.(1989)Clough, Kneizys, \& Davies]{clough1989line}
Clough, S.~A., Kneizys, F.~X., \& Davies, R.~W., 1989.
\newblock Line shape and the water vapor continuum, {\it Atmospheric
  research\/}, {\bf 23}, 229--241.

\bibitem[Clough et~al.(2005)Clough, Shephard, Mlawer, Delamere, Iacono,
  Cady-Pereira, Boukabara, \& Brown]{clough2005atmospheric}
Clough, S.~A., Shephard, M.~W., Mlawer, E.~J., Delamere, J.~S., Iacono, M.~J.,
  Cady-Pereira, K., Boukabara, S., \& Brown, P.~D., 2005.
\newblock Atmospheric radiative transfer modeling: A summary of the aer codes,
  {\it Journal of Quantitative Spectroscopy and Radiative Transfer\/}, {\bf
  91}, 233--244.

\bibitem[Cole et~al.(2023)Cole, Tran, Hoghooghi, \& Rieker]{Cole2023}
Cole, R.~K., Tran, H., Hoghooghi, N., \& Rieker, G.~B., 2023.
\newblock {Temperature-dependent CO2 Line Mixing Models using Dual Frequency
  Comb Absorption and Phase Spectroscopy up to 25 bar and 1000 K}, {\it Journal
  of Quantitative Spectroscopy and Radiative Transfer\/}, p. 108488.

\bibitem[Delahaye et~al.(2021)Delahaye, Armante, Scott, Jacquinet-Husson,
  Ch{\'e}din, Cr{\'e}peau, Crevoisier, Douet, Perrin, Barbe,
  et~al.]{delahaye20212020}
Delahaye, T., Armante, R., Scott, N., Jacquinet-Husson, N., Ch{\'e}din, A.,
  Cr{\'e}peau, L., Crevoisier, C., Douet, V., Perrin, A., Barbe, A., et~al.,
  2021.
\newblock The 2020 edition of the geisa spectroscopic database, {\it Journal of
  Molecular Spectroscopy\/}, {\bf 380}, 111510.

\bibitem[{Fortney} et~al.(2005){Fortney}, {Marley}, {Lodders}, {Saumon}, \&
  {Freedman}]{Fortney2005ApJ}
{Fortney}, J.~J., {Marley}, M.~S., {Lodders}, K., {Saumon}, D., \& {Freedman},
  R., 2005.
\newblock {Comparative Planetary Atmospheres: Models of TrES-1 and HD 209458b},
  {\it \apjl\/}, {\bf 627}(1), L69--L72.

\bibitem[Fortney et~al.(2019)Fortney, Robinson, Domagal-Goldman, Del~Genio,
  Gordon, Gharib-Nezhad, Lewis, Sousa-Silva, Airapetian, Drouin,
  et~al.]{fortney2019need}
Fortney, J.~J., Robinson, T.~D., Domagal-Goldman, S., Del~Genio, A.~D., Gordon,
  I.~E., Gharib-Nezhad, E., Lewis, N., Sousa-Silva, C., Airapetian, V., Drouin,
  B., et~al., 2019.
\newblock The need for laboratory measurements and ab initio studies to aid
  understanding of exoplanetary atmospheres, {\it arXiv preprint
  arXiv:1905.07064\/}.

\bibitem[Freedman et~al.(2014)Freedman, Lustig-Yaeger, Fortney, Lupu, Marley,
  \& Lodders]{14FrLuFo}
Freedman, R.~S., Lustig-Yaeger, J., Fortney, J.~J., Lupu, R.~E., Marley, M.~S.,
  \& Lodders, K., 2014.
\newblock Gaseous mean opacities for giant planet and ultracool dwarf
  atmospheres over a range of metallicities and temperatures, {\it The
  Astrophysical Journal Supplement Series\/}, {\bf 214}, 25.

\bibitem[Fu et~al.(1996)Fu, Borysow, \& Moraldi]{1996Fu}
Fu, Y., Borysow, A., \& Moraldi, M., 1996.
\newblock High-frequency wings of rototranslational raman spectra of gaseous
  nitrogen, {\it Phys. Rev. A\/}, {\bf 53}, 201--205.

\bibitem[Gamache \& Laraia(2009)]{Gamache2009}
Gamache, R.~R. \& Laraia, A.~L., 2009.
\newblock {N$_2$-}, {O$_2$-}, and air-broadened half-widths, their temperature
  dependence, and line shifts for the rotation band of {H$_2$$^{16}$O}, {\it
  Journal of Molecular Spectroscopy\/}, {\bf 257}, 116--127.

\bibitem[Gamache et~al.(2019)Gamache, Rey, Vispoel, \& Tyuterev]{GAMACHE2019}
Gamache, R.~R., Rey, M., Vispoel, B., \& Tyuterev, V.~G., 2019.
\newblock Reduced matrix elements for collisionally induced transitions of
  12ch4, {\it Journal of Quantitative Spectroscopy and Radiative Transfer\/},
  {\bf 235}, 31--39.

\bibitem[Gharib-Nezhad \& Line(2019)]{GharibNezhad2019steam}
Gharib-Nezhad, E. \& Line, M.~R., 2019.
\newblock The influence of {H$_2$O} pressure broadening in high-metallicity
  exoplanet atmospheres, {\it The Astrophysical Journal\/}, {\bf 872}(1), 27.

\bibitem[Gharib-Nezhad et~al.(2021{\natexlab{a}})Gharib-Nezhad, Iyer, Line,
  Freedman, Marley, \& Batalha]{21GhIyLi}
Gharib-Nezhad, E., Iyer, A.~R., Line, M.~R., Freedman, R.~S., Marley, M.~S., \&
  Batalha, N.~E., 2021{\natexlab{a}}.
\newblock {EXOPLINES: M}olecular absorption cross-section database for brown
  dwarf and giant exoplanet atmospheres, {\it The Astrophysical Journal
  Supplement Series\/}, {\bf 254}, 34.

\bibitem[Gharib-Nezhad et~al.(2021{\natexlab{b}})Gharib-Nezhad, Marley,
  Batalha, Visscher, Freedman, \& Lupu]{GharibNezhad2021Li}
Gharib-Nezhad, E., Marley, M.~S., Batalha, N.~E., Visscher, C., Freedman,
  R.~S., \& Lupu, R.~E., 2021{\natexlab{b}}.
\newblock Following the lithium: Tracing {Li}-bearing molecules across age,
  mass, and gravity in brown dwarfs, {\it The Astrophysical Journal\/}, {\bf
  919}(1), 21.

\bibitem[Gordon et~al.(2022)Gordon, Rothman, Hargreaves, Hashemi, Karlovets,
  Skinner, Conway, Hill, Kochanov, Tan, Wcisło, Finenko, Nelson, Bernath,
  Birk, Boudon, Campargue, Chance, Coustenis, Drouin, Flaud, Gamache, Hodges,
  Jacquemart, Mlawer, Nikitin, Perevalov, Rotger, Tennyson, Toon, Tran,
  Tyuterev, Adkins, Baker, Barbe, Canè, Császár, Dudaryonok, Egorov,
  Fleisher, Fleurbaey, Foltynowicz, Furtenbacher, Harrison, Hartmann, Horneman,
  Huang, Karman, Karns, Kassi, Kleiner, Kofman, Kwabia–Tchana, Lavrentieva,
  Lee, Long, Lukashevskaya, Lyulin, Makhnev, Matt, Massie, Melosso,
  Mikhailenko, Mondelain, Müller, Naumenko, Perrin, Polyansky, Raddaoui,
  Raston, Reed, Rey, Richard, Tóbiás, Sadiek, Schwenke, Starikova, Sung,
  Tamassia, Tashkun, {Vander Auwera}, Vasilenko, Vigasin, Villanueva, Vispoel,
  Wagner, Yachmenev, \& Yurchenko]{HITRAN2020}
Gordon, I., Rothman, L., Hargreaves, R., Hashemi, R., Karlovets, E., Skinner,
  F., Conway, E., Hill, C., Kochanov, R., Tan, Y., Wcisło, P., Finenko, A.,
  Nelson, K., Bernath, P., Birk, M., Boudon, V., Campargue, A., Chance, K.,
  Coustenis, A., Drouin, B., Flaud, J., Gamache, R., Hodges, J., Jacquemart,
  D., Mlawer, E., Nikitin, A., Perevalov, V., Rotger, M., Tennyson, J., Toon,
  G., Tran, H., Tyuterev, V., Adkins, E., Baker, A., Barbe, A., Canè, E.,
  Császár, A., Dudaryonok, A., Egorov, O., Fleisher, A., Fleurbaey, H.,
  Foltynowicz, A., Furtenbacher, T., Harrison, J., Hartmann, J., Horneman, V.,
  Huang, X., Karman, T., Karns, J., Kassi, S., Kleiner, I., Kofman, V.,
  Kwabia–Tchana, F., Lavrentieva, N., Lee, T., Long, D., Lukashevskaya, A.,
  Lyulin, O., Makhnev, V., Matt, W., Massie, S., Melosso, M., Mikhailenko, S.,
  Mondelain, D., Müller, H., Naumenko, O., Perrin, A., Polyansky, O.,
  Raddaoui, E., Raston, P., Reed, Z., Rey, M., Richard, C., Tóbiás, R.,
  Sadiek, I., Schwenke, D., Starikova, E., Sung, K., Tamassia, F., Tashkun, S.,
  {Vander Auwera}, J., Vasilenko, I., Vigasin, A., Villanueva, G., Vispoel, B.,
  Wagner, G., Yachmenev, A., \& Yurchenko, S., 2022.
\newblock The hitran2020 molecular spectroscopic database, {\it Journal of
  Quantitative Spectroscopy and Radiative Transfer\/}, {\bf 277}, 107949.

\bibitem[Grimm \& Heng(2015)]{15GrHe}
Grimm, S.~L. \& Heng, K., 2015.
\newblock Helios-k: An ultrafast, open-source opacity calculator for radiative
  transfer, {\it The Astrophysical Journal\/}, {\bf 808}, 182.

\bibitem[Grimm et~al.(2021)Grimm, Malik, Kitzmann, Guzm\'{a}n-Mesa,
  Hoeijmakers, Fisher, {a}o M.~Mendon\c{c}a, Yurchenko, Tennyson, Kurucz, \&
  Heng]{jt819}
Grimm, S.~L., Malik, M., Kitzmann, D., Guzm\'{a}n-Mesa, A., Hoeijmakers, H.~J.,
  Fisher, C., {a}o M.~Mendon\c{c}a, J., Yurchenko, S.~N., Tennyson, J., Kurucz,
  R.~L., \& Heng, K., 2021.
\newblock {HELIOS-K 2.0 and an Open-source Opacity Database for Exoplanetary
  Atmospheres}, {\it Astrophys. J. Suppl. Ser.\/}, {\bf 253}, 30.

\bibitem[Gunson et~al.(1996)Gunson, Abbas, Abrams, Allen, Brown, Brown, Chang,
  Goldman, Irion, Lowes, et~al.]{gunson1996atmospheric}
Gunson, M.~R., Abbas, M.~M., Abrams, M.~C., Allen, M., Brown, L.~R., Brown,
  T.~L., Chang, A.~Y., Goldman, A., Irion, F.~W., Lowes, L.~L., et~al., 1996.
\newblock The atmospheric trace molecule spectroscopy (atmos) experiment:
  Deployment on the atlas space shuttle missions, {\it Geophysical Research
  Letters\/}, {\bf 23}, 2333--2336.

\bibitem[Hartmann et~al.(2002)Hartmann, Boulet, \& Brodbeck]{Hartmann2002}
Hartmann, J.-M., Boulet, C., \& Brodbeck, C., 2002.
\newblock {A far wing lineshape for H 2 broadened CH 4 infrared transitions},
  {\it Journal of Quantitative Spectroscopy and Radiative Transfer\/}, {\bf
  72}, 117--122.

\bibitem[Hartmann et~al.(2008)Hartmann, Boulet, \& Robert]{2008Hartmann}
Hartmann, J.-M., Boulet, C., \& Robert, D., 2008.
\newblock {\it Collisional Effects on Molecular Spectra: Laboratory Experiments
  and Models, Consequences for Applications\/}, Elsevier Science.

\bibitem[Hashemi et~al.(2020)Hashemi, Gordon, Tran, Kochanov, Karlovets, Tan,
  Lamouroux, Ngo, \& Rothman]{Hashemi_HITRANCO2}
Hashemi, R., Gordon, I.~E., Tran, H., Kochanov, R.~V., Karlovets, E.~V., Tan,
  Y., Lamouroux, J., Ngo, N.~H., \& Rothman, L.~S., 2020.
\newblock {Revising the line-shape parameters for air- and self-broadened
  CO$_2$ lines toward a sub-percent accuracy level}, {\it J. Quant. Spectrosc.
  Radiat. Transf.\/}, {\bf 256}, 107283.

\bibitem[Hashemi et~al.(2021)Hashemi, Gordon, Adkins, Hodges, Long, Birk, Loos,
  Boone, Fleisher, Predoi-Cross, \& Rothman]{Hashemi_HITRANN2OCO}
Hashemi, R., Gordon, I.~E., Adkins, E.~M., Hodges, J.~T., Long, D.~A., Birk,
  M., Loos, J., Boone, C.~D., Fleisher, A.~J., Predoi-Cross, A., \& Rothman,
  L.~S., 2021.
\newblock Improvement of the spectroscopic parameters of the air- and
  self-broadened n2o and co lines for the hitran2020 database applications,
  {\it J. Quant. Spectrosc. Radiat. Transf.\/}, {\bf 271}, 107735.

\bibitem[Hedges \& Madhusudhan(2016)]{16HeMaxx}
Hedges, C. \& Madhusudhan, N., 2016.
\newblock Effect of pressure broadening on molecular absorption cross sections
  in exoplanetary atmospheres, {\it Mon. Not. R. Astron. Soc.\/}, {\bf 458},
  1427--1449.

\bibitem[Hill et~al.(2013)Hill, Yurchenko, \& Tennyson]{13HiYuTe}
Hill, C., Yurchenko, S.~N., \& Tennyson, J., 2013.
\newblock Temperature-dependent molecular absorption cross sections for
  exoplanets and other atmospheres, {\it Icarus\/}, {\bf 226}, 1673--1677.

\bibitem[{Hood} et~al.(2023){Hood}, {Fortney}, {Line}, \& {Faherty}]{Hood2023}
{Hood}, C.~E., {Fortney}, J.~J., {Line}, M.~R., \& {Faherty}, J.~K., 2023.
\newblock {Brown Dwarf Retrievals on FIRE!: Atmospheric Constraints and Lessons
  Learned from High Signal-to-Noise Medium Resolution Spectroscopy of a T9
  Dwarf}, {\it arXiv e-prints\/}, p. arXiv:2303.04885.

\bibitem[Huml\'{i}\v{c}ek({1979})]{79Humlic}
Huml\'{i}\v{c}ek, J., {1979}.
\newblock Efficient method for evaluation of the complex probability function -
  {V}oigt function and its derivatives, {\it J. Quant. Spectrosc. Radiat.
  Transf.\/}, {\bf {21}}, 309--313.

\bibitem[Hussey et~al.(1975)Hussey, Dufty, \& Hooper]{Hussey1975Kinetic}
Hussey, T., Dufty, J.~W., \& Hooper, C.~F., 1975.
\newblock Kinetic theory of spectral line broadening, {\it Phys. Rev. A\/},
  {\bf 12}, 1084--1093.

\bibitem[Iyer et~al.(2023)Iyer, Line, Muirhead, Fortney, \&
  Gharib-Nezhad]{iyer2023sphinx}
Iyer, A.~R., Line, M.~R., Muirhead, P.~S., Fortney, J.~J., \& Gharib-Nezhad,
  E., 2023.
\newblock The sphinx m-dwarf spectral grid. i. benchmarking new model
  atmospheres to derive fundamental m-dwarf properties, {\it The Astrophysical
  Journal\/}, {\bf 944}(1), 41.

\bibitem[{JWST Transiting Exoplanet Community Early Release Science Team}
  et~al.(2023){JWST Transiting Exoplanet Community Early Release Science Team},
  {Ahrer}, {Alderson}, {Batalha}, {Batalha}, {Bean}, {Beatty}, {Bell},
  {Benneke}, {Berta-Thompson}, {Carter}, {Crossfield}, {Espinoza}, {Feinstein},
  {Fortney}, {Gibson}, {Goyal}, {Kempton}, {Kirk}, {Kreidberg},
  {L{\'o}pez-Morales}, {Line}, {Lothringer}, {Moran}, {Mukherjee}, {Ohno},
  {Parmentier}, {Piaulet}, {Rustamkulov}, {Schlawin}, {Sing}, {Stevenson},
  {Wakeford}, {Allen}, {Birkmann}, {Brande}, {Crouzet}, {Cubillos}, {Damiano},
  {D{\'e}sert}, {Gao}, {Harrington}, {Hu}, {Kendrew}, {Knutson}, {Lagage},
  {Leconte}, {Lendl}, {MacDonald}, {May}, {Miguel}, {Molaverdikhani}, {Moses},
  {Murray}, {Nehring}, {Nikolov}, {Petit dit de la Roche}, {Radica}, {Roy},
  {Stassun}, {Taylor}, {Waalkes}, {Wachiraphan}, {Welbanks}, {Wheatley},
  {Aggarwal}, {Alam}, {Banerjee}, {Barstow}, {Blecic}, {Casewell}, {Changeat},
  {Chubb}, {Col{\'o}n}, {Coulombe}, {Daylan}, {de Val-Borro}, {Decin}, {Dos
  Santos}, {Flagg}, {France}, {Fu}, {Garc{\'\i}a Mu{\~n}oz}, {Gizis},
  {Glidden}, {Grant}, {Heng}, {Henning}, {Hong}, {Inglis}, {Iro}, {Kataria},
  {Komacek}, {Krick}, {Lee}, {Lewis}, {Lillo-Box}, {Lustig-Yaeger}, {Mancini},
  {Mandell}, {Mansfield}, {Marley}, {Mikal-Evans}, {Morello}, {Nixon}, {Ortiz
  Ceballos}, {Piette}, {Powell}, {Rackham}, {Ramos-Rosado}, {Rauscher},
  {Redfield}, {Rogers}, {Roman}, {Roudier}, {Scarsdale}, {Shkolnik},
  {Southworth}, {Spake}, {Steinrueck}, {Tan}, {Teske}, {Tremblin}, {Tsai},
  {Tucker}, {Turner}, {Valenti}, {Venot}, {Waldmann}, {Wallack}, {Zhang}, \&
  {Zieba}]{jwst2023identification}
{JWST Transiting Exoplanet Community Early Release Science Team}, {Ahrer},
  E.-M., {Alderson}, L., {Batalha}, N.~M., {Batalha}, N.~E., {Bean}, J.~L.,
  {Beatty}, T.~G., {Bell}, T.~J., {Benneke}, B., {Berta-Thompson}, Z.~K.,
  {Carter}, A.~L., {Crossfield}, I. J.~M., {Espinoza}, N., {Feinstein}, A.~D.,
  {Fortney}, J.~J., {Gibson}, N.~P., {Goyal}, J.~M., {Kempton}, E. M.~R.,
  {Kirk}, J., {Kreidberg}, L., {L{\'o}pez-Morales}, M., {Line}, M.~R.,
  {Lothringer}, J.~D., {Moran}, S.~E., {Mukherjee}, S., {Ohno}, K.,
  {Parmentier}, V., {Piaulet}, C., {Rustamkulov}, Z., {Schlawin}, E., {Sing},
  D.~K., {Stevenson}, K.~B., {Wakeford}, H.~R., {Allen}, N.~H., {Birkmann},
  S.~M., {Brande}, J., {Crouzet}, N., {Cubillos}, P.~E., {Damiano}, M.,
  {D{\'e}sert}, J.-M., {Gao}, P., {Harrington}, J., {Hu}, R., {Kendrew}, S.,
  {Knutson}, H.~A., {Lagage}, P.-O., {Leconte}, J., {Lendl}, M., {MacDonald},
  R.~J., {May}, E.~M., {Miguel}, Y., {Molaverdikhani}, K., {Moses}, J.~I.,
  {Murray}, C.~A., {Nehring}, M., {Nikolov}, N.~K., {Petit dit de la Roche},
  D.~J.~M., {Radica}, M., {Roy}, P.-A., {Stassun}, K.~G., {Taylor}, J.,
  {Waalkes}, W.~C., {Wachiraphan}, P., {Welbanks}, L., {Wheatley}, P.~J.,
  {Aggarwal}, K., {Alam}, M.~K., {Banerjee}, A., {Barstow}, J.~K., {Blecic},
  J., {Casewell}, S.~L., {Changeat}, Q., {Chubb}, K.~L., {Col{\'o}n}, K.~D.,
  {Coulombe}, L.-P., {Daylan}, T., {de Val-Borro}, M., {Decin}, L., {Dos
  Santos}, L.~A., {Flagg}, L., {France}, K., {Fu}, G., {Garc{\'\i}a Mu{\~n}oz},
  A., {Gizis}, J.~E., {Glidden}, A., {Grant}, D., {Heng}, K., {Henning}, T.,
  {Hong}, Y.-C., {Inglis}, J., {Iro}, N., {Kataria}, T., {Komacek}, T.~D.,
  {Krick}, J.~E., {Lee}, E. K.~H., {Lewis}, N.~K., {Lillo-Box}, J.,
  {Lustig-Yaeger}, J., {Mancini}, L., {Mandell}, A.~M., {Mansfield}, M.,
  {Marley}, M.~S., {Mikal-Evans}, T., {Morello}, G., {Nixon}, M.~C., {Ortiz
  Ceballos}, K., {Piette}, A. A.~A., {Powell}, D., {Rackham}, B.~V.,
  {Ramos-Rosado}, L., {Rauscher}, E., {Redfield}, S., {Rogers}, L.~K., {Roman},
  M.~T., {Roudier}, G.~M., {Scarsdale}, N., {Shkolnik}, E.~L., {Southworth},
  J., {Spake}, J.~J., {Steinrueck}, M.~E., {Tan}, X., {Teske}, J.~K.,
  {Tremblin}, P., {Tsai}, S.-M., {Tucker}, G.~S., {Turner}, J.~D., {Valenti},
  J.~A., {Venot}, O., {Waldmann}, I.~P., {Wallack}, N.~L., {Zhang}, X., \&
  {Zieba}, S., 2023.
\newblock {Identification of carbon dioxide in an exoplanet atmosphere}, {\it
  \nat\/}, {\bf 614}, 649--652.

\bibitem[{Kochanov} et~al.(2016){Kochanov}, {Gordon}, {Rothman}, {Wcis{\l}o},
  {Hill}, \& {Wilzewski}]{HAPI-REF}
{Kochanov}, R.~V., {Gordon}, I.~E., {Rothman}, L.~S., {Wcis{\l}o}, P., {Hill},
  C., \& {Wilzewski}, J.~S., 2016.
\newblock {HITRAN Application Programming Interface (HAPI): A comprehensive
  approach to working with spectroscopic data}, {\it \jqsrt\/}, {\bf 177},
  15--30.

\bibitem[Lamouroux et~al.(2010)Lamouroux, Tran, Laraia, Gamache, Rothman,
  Gordon, \& Hartmann]{LAMOUROUX2010}
Lamouroux, J., Tran, H., Laraia, A.~L., Gamache, R.~R., Rothman, L.~S., Gordon,
  I.~E., \& Hartmann, J.-M., 2010.
\newblock {Updated database plus software for line-mixing in CO2 infrared
  spectra and their test using laboratory spectra in the 1.5–2.3 $\mu$m
  region}, {\it Journal of Quantitative Spectroscopy and Radiative Transfer\/},
  {\bf 111}, 2321--2331, XVIth Symposium on High Resolution Molecular
  Spectroscopy (HighRus-2009).

\bibitem[Lamouroux et~al.(2014)Lamouroux, Gamache, \& Schwenke]{LAMOUROUX2014}
Lamouroux, J., Gamache, R., \& Schwenke, D., 2014.
\newblock Determination of the reduced matrix elements using accurate ab initio
  wavefunctions: Formalism and its application to the vibrational ground state
  (000) of h216o, {\it Journal of Quantitative Spectroscopy and Radiative
  Transfer\/}, {\bf 148}, 49--57.

\bibitem[Letchworth \& Benner(2007)]{letchworth2007rapid}
Letchworth, K.~L. \& Benner, D.~C., 2007.
\newblock Rapid and accurate calculation of the voigt function, {\it Journal of
  Quantitative Spectroscopy and Radiative Transfer\/}, {\bf 107}(1), 173--192.

\bibitem[L\'{e}vy et~al.(1992)L\'{e}vy, Lacome, \& Chackerian]{1992Levy}
L\'{e}vy, A., Lacome, N., \& Chackerian, C., 1992.
\newblock Collisional line mixing, in {\em Spectroscopy of the Earth's
  Atmosphere and Interstellar Medium\/}, pp. 261--337, eds Rao, K.~N. \& Weber,
  A., Academic Press.

\bibitem[{Lothringer} et~al.(2018){Lothringer}, {Barman}, \&
  {Koskinen}]{Lothringer2018ApJ}
{Lothringer}, J.~D., {Barman}, T., \& {Koskinen}, T., 2018.
\newblock {Extremely Irradiated Hot Jupiters: Non-oxide Inversions, H$^{-}$
  Opacity, and Thermal Dissociation of Molecules}, {\it \apj\/}, {\bf 866}, 27.

\bibitem[Ma \& Tipping(1990{\natexlab{a}})]{Ma1990Water-millimeter-JCP}
Ma, Q. \& Tipping, R.~H., 1990{\natexlab{a}}.
\newblock Water vapor continuum in the millimeter spectral region, {\it The
  Journal of Chemical Physics\/}, {\bf 93}(9), 6127--6139.

\bibitem[Ma \& Tipping(1990{\natexlab{b}})]{ma1990atmospheric}
Ma, Q. \& Tipping, R.~H., 1990{\natexlab{b}}.
\newblock The atmospheric water continuum in the infrared: extension of the
  statistical theory of rosenkranz, {\it The Journal of chemical physics\/},
  {\bf 93}, 7066--7075.

\bibitem[Ma \& Tipping(1990{\natexlab{c}})]{ma1990water}
Ma, Q. \& Tipping, R.~H., 1990{\natexlab{c}}.
\newblock Water vapor continuum in the millimeter spectral region, {\it The
  Journal of chemical physics\/}, {\bf 93}, 6127--6139.

\bibitem[Ma \& Tipping(1991)]{ma1991far}
Ma, Q. \& Tipping, R.~H., 1991.
\newblock A far wing line shape theory and its application to the water
  continuum absorption in the infrared region. i, {\it The Journal of chemical
  physics\/}, {\bf 95}, 6290--6301.

\bibitem[Ma \& Tipping(1992)]{ma1992far}
Ma, Q. \& Tipping, R.~H., 1992.
\newblock A far wing line shape theory and its application to the water
  vibrational bands (ii), {\it The Journal of chemical physics\/}, {\bf 96},
  8655--8663.

\bibitem[Ma \& Tipping(1994)]{ma1994detailed}
Ma, Q. \& Tipping, R.~H., 1994.
\newblock The detailed balance requirement and general empirical formalisms for
  continuum absorption, {\it Journal of Quantitative Spectroscopy and Radiative
  Transfer\/}, {\bf 51}, 751--757.

\bibitem[Ma et~al.(1999)Ma, Tipping, Boulet, \& Bouanich]{ma1999theoretical}
Ma, Q., Tipping, R.~H., Boulet, C., \& Bouanich, J.-P., 1999.
\newblock Theoretical far-wing line shape and absorption for high-temperature
  co 2, {\it Applied optics\/}, {\bf 38}, 599--604.

\bibitem[MacDonald(2019)]{MacDonaldPhD}
MacDonald, R.~J., 2019.
\newblock Revealing the nature of exoplanetary atmospheres, {\it PhD thesis,
  University of Cambridge, UK\/}.

\bibitem[McClatchey et~al.(1973)McClatchey, Benedict, Clough, Burch, Calfee,
  Fox, Rothman, \& Garing]{AFGL1973}
McClatchey, R., Benedict, W., Clough, S., Burch, D., Calfee, R., Fox, K.,
  Rothman, L., \& Garing, J., 1973.
\newblock {AFCRL Atmospheric Absorption Line Parameters Compilation}, {\it
  Environmental Research Papers\/}, {\bf 434}, 1--86.

\bibitem[Meadows \& Crisp(1996)]{SMART}
Meadows, V.~S. \& Crisp, D., 1996.
\newblock {Ground-based near-infrared observations of the Venus nightside: The
  thermal structure and water abundance near the surface}, {\it Journal of
  Geophysical Research\/}, {\bf 101}, 4595.

\bibitem[{Miles} et~al.(2022){Miles}, {Biller}, {Patapis}, {Worthen},
  {Rickman}, {Hoch}, {Skemer}, {Perrin}, {Whiteford}, {Chen}, {Mukherjee},
  {Morley}, {Moran}, {Bonnefoy}, {Petrus}, {Carter}, {Choquet}, {Hinkley},
  {Ward-Duong}, {Leisenring}, {Millar-Blanchaer}, {Pueyo}, {Ray},
  {Stapelfeldt}, {Stone}, {Wang}, {Absil}, {Balmer}, {Boccaletti}, {Bonavita},
  {Booth}, {Bowler}, {Chauvin}, {Christiaens}, {Currie}, {Danielski},
  {Fortney}, {Girard}, {Greenbaum}, {Henning}, {Hines}, {Janson}, {Kalas},
  {Kammerer}, {Kenworthy}, {Kervella}, {Lagage}, {Lew}, {Liu}, {Macintosh},
  {Marino}, {Marley}, {Marois}, {Matthews}, {Matthews}, {Mawet}, {McElwain},
  {Metchev}, {Meyer}, {Molliere}, {Pantin}, {Rebollido}, {Ren}, {Vasist},
  {Wyatt}, {Zhou}, {Briesemeister}, {Bryan}, {Calissendorff}, {Catalloube},
  {Cugno}, {De Furio}, {Dupuy}, {Factor}, {Faherty}, {Fitzgerald}, {Franson},
  {Gonzales}, {Hood}, {Howe}, {Kraus}, {Kuzuhara}, {Lawson}, {Lazzoni}, {Liu},
  {Llop-Sayson}, {Lloyd}, {Martinez}, {Mazoyer}, {Quanz}, {Adams Redai},
  {Samland}, {Schlieder}, {Tamura}, {Tan}, {Uyama}, {Vigan}, {Vos}, {Wagner},
  {Wolff}, {Ygouf}, {Zhang}, \& {Zhang}]{Miles2022}
{Miles}, B.~E., {Biller}, B.~A., {Patapis}, P., {Worthen}, K., {Rickman}, E.,
  {Hoch}, K. K.~W., {Skemer}, A., {Perrin}, M.~D., {Whiteford}, N., {Chen},
  C.~H., {Mukherjee}, S., {Morley}, C.~V., {Moran}, S.~E., {Bonnefoy}, M.,
  {Petrus}, S., {Carter}, A.~L., {Choquet}, E., {Hinkley}, S., {Ward-Duong},
  K., {Leisenring}, J.~M., {Millar-Blanchaer}, M.~A., {Pueyo}, L., {Ray}, S.,
  {Stapelfeldt}, K.~R., {Stone}, J.~M., {Wang}, J.~J., {Absil}, O., {Balmer},
  W.~O., {Boccaletti}, A., {Bonavita}, M., {Booth}, M., {Bowler}, B.~P.,
  {Chauvin}, G., {Christiaens}, V., {Currie}, T., {Danielski}, C., {Fortney},
  J.~J., {Girard}, J.~H., {Greenbaum}, A.~Z., {Henning}, T., {Hines}, D.~C.,
  {Janson}, M., {Kalas}, P., {Kammerer}, J., {Kenworthy}, M.~A., {Kervella},
  P., {Lagage}, P.-O., {Lew}, B. W.~P., {Liu}, M.~C., {Macintosh}, B.,
  {Marino}, S., {Marley}, M.~S., {Marois}, C., {Matthews}, E.~C., {Matthews},
  B.~C., {Mawet}, D., {McElwain}, M.~W., {Metchev}, S., {Meyer}, M.~R.,
  {Molliere}, P., {Pantin}, E., {Rebollido}, A. Q.~I., {Ren}, B.~B., {Vasist},
  M., {Wyatt}, M.~C., {Zhou}, Y., {Briesemeister}, Z.~W., {Bryan}, M.~L.,
  {Calissendorff}, P., {Catalloube}, F., {Cugno}, G., {De Furio}, M., {Dupuy},
  T.~J., {Factor}, S.~M., {Faherty}, J.~K., {Fitzgerald}, M.~P., {Franson}, K.,
  {Gonzales}, E.~C., {Hood}, C.~E., {Howe}, A.~R., {Kraus}, A.~L., {Kuzuhara},
  M., {Lawson}, K., {Lazzoni}, C., {Liu}, P., {Llop-Sayson}, J., {Lloyd},
  J.~P., {Martinez}, R.~A., {Mazoyer}, J., {Quanz}, S.~P., {Adams Redai}, J.,
  {Samland}, M., {Schlieder}, J.~E., {Tamura}, M., {Tan}, X., {Uyama}, T.,
  {Vigan}, A., {Vos}, J.~M., {Wagner}, K., {Wolff}, S.~G., {Ygouf}, M.,
  {Zhang}, K., \& {Zhang}, Z., 2022.
\newblock {The JWST Early Release Science Program for Direct Observations of
  Exoplanetary Systems II: A 1 to 20 Micron Spectrum of the Planetary-Mass
  Companion VHS 1256-1257 b}, {\it arXiv e-prints\/}, p. arXiv:2209.00620.

\bibitem[Mlawer et~al.(2012)Mlawer, Payne, Moncet, Delamere, Alvarado, \&
  Tobin]{mlawer2012development}
Mlawer, E.~J., Payne, V.~H., Moncet, J.-L., Delamere, J.~S., Alvarado, M.~J.,
  \& Tobin, D.~C., 2012.
\newblock Development and recent evaluation of the mt\_ckd model of continuum
  absorption, {\it Philosophical Transactions of the Royal Society A:
  Mathematical, Physical and Engineering Sciences\/}, {\bf 370}, 2520--2556.

\bibitem[{Molli{\`e}re} et~al.(2019){Molli{\`e}re}, {Wardenier}, {van Boekel},
  {Henning}, {Molaverdikhani}, \& {Snellen}]{molliere2019}
{Molli{\`e}re}, P., {Wardenier}, J.~P., {van Boekel}, R., {Henning}, T.,
  {Molaverdikhani}, K., \& {Snellen}, I.~A.~G., 2019.
\newblock {petitRADTRANS. A Python radiative transfer package for exoplanet
  characterization and retrieval}, {\it \aap\/}, {\bf 627}, A67.

\bibitem[{Mukherjee} et~al.(2023){Mukherjee}, {Batalha}, {Fortney}, \&
  {Marley}]{mukherjee2023}
{Mukherjee}, S., {Batalha}, N.~E., {Fortney}, J.~J., \& {Marley}, M.~S., 2023.
\newblock {PICASO 3.0: A One-dimensional Climate Model for Giant Planets and
  Brown Dwarfs}, {\it \apj\/}, {\bf 942}, 71.

\bibitem[Ngo et~al.(2013)Ngo, Lisak, Tran, \& Hartmann]{13NgLiTr}
Ngo, N., Lisak, D., Tran, H., \& Hartmann, J.-M., 2013.
\newblock An isolated line-shape model to go beyond the {Voigt} profile in
  spectroscopic databases and radiative transfer codes, {\it Journal of
  Quantitative Spectroscopy and Radiative Transfer\/}, {\bf 129}, 89--100.

\bibitem[Ngo et~al.(2017)Ngo, Lin, Hodges, \& Tran]{17NgLiHo}
Ngo, N., Lin, H., Hodges, J., \& Tran, H., 2017.
\newblock Spectral shapes of rovibrational lines of {CO} broadened by {He, Ar,
  Kr and SF$_6$}: A test case of the {Hartmann-Tran} profile, {\it Journal of
  Quantitative Spectroscopy and Radiative Transfer\/}, {\bf 203}, 325--333.

\bibitem[Niraula et~al.(2022)Niraula, de~Wit, Gordon, Hargreaves, Sousa-Silva,
  \& Kochanov]{22NiWiGo}
Niraula, P., de~Wit, J., Gordon, I.~E., Hargreaves, R.~J., Sousa-Silva, C., \&
  Kochanov, R.~V., 2022.
\newblock The impending opacity challenge in exoplanet atmospheric
  characterization, {\it Nature Astronomy\/}, {\bf 6}, 1287--1295.

\bibitem[Niraula et~al.(2023)Niraula, de~Wit, Gordon, Hargreaves, \&
  Sousa-Silva]{Niraula2023}
Niraula, P., de~Wit, J., Gordon, I.~E., Hargreaves, R.~J., \& Sousa-Silva, C.,
  2023.
\newblock {Origin and Extent of the Opacity Challenge for Atmospheric
  Retrievals of WASP-39 b}, {\it The Astrophysical Journal Letters\/}, {\bf
  950}(2), L17.

\bibitem[{Olivero} \& {Longbothum}(1977)]{Olivero1977}
{Olivero}, J. \& {Longbothum}, R., 1977.
\newblock {Empirical fits to the Voigt line width: A brief review}, {\it
  \jqsrt\/}, {\bf 17}, 233--236.

\bibitem[Pieroni et~al.(2001)Pieroni, Hartmann, Camy-Peyret, Jeseck, \&
  Payan]{PIERONI2001}
Pieroni, D., Hartmann, J.-M., Camy-Peyret, C., Jeseck, P., \& Payan, S., 2001.
\newblock Influence of line mixing on absorption by ch4 in atmospheric
  balloon-borne spectra near 3.3 $\mu$m, {\it Journal of Quantitative
  Spectroscopy and Radiative Transfer\/}, {\bf 68}(2), 117--133.

\bibitem[Pollack et~al.(1993)Pollack, Dalton, Grinspoon, Wattson, Freedman,
  Crisp, Allen, Bezard, DeBergh, Giver, Ma, \& Tipping]{93PoDaGr}
Pollack, J.~B., Dalton, J., Grinspoon, D., Wattson, R.~B., Freedman, R., Crisp,
  D., Allen, D.~A., Bezard, B., DeBergh, C., Giver, L.~P., Ma, Q., \& Tipping,
  R., 1993.
\newblock Near-infrared light from {Venus'} nightside: A spectroscopic
  analysis, {\it Icarus\/}, {\bf 103}(1), 1--42.

\bibitem[Ren et~al.(2023)Ren, Han, Modest, Fateev, \& Clausen]{Ren2023}
Ren, T., Han, Y., Modest, M.~F., Fateev, A., \& Clausen, S., 2023.
\newblock Evaluation of spectral line mixing models and the effects on high
  pressure radiative heat transfer calculations, {\it Journal of Quantitative
  Spectroscopy and Radiative Transfer\/}, {\bf 302}, 108555.

\bibitem[Rey et~al.(2016)Rey, Nikitin, Babikov, \& Tyuterev]{TheoReTS}
Rey, M., Nikitin, A.~V., Babikov, Y.~L., \& Tyuterev, V.~G., 2016.
\newblock {TheoReTS} - an information system for theoretical spectra based on
  variational predictions from molecular potential energy and dipole moment
  surfaces, {\it Journal of Molecular Spectroscopy\/}, {\bf 327}, 138 -- 158,
  New Visions of Spectroscopic Databases, Volume II.

\bibitem[Rothman et~al.(2010)Rothman, Gordon, Barber, Dothe, Gamache, Goldman,
  Perevalov, Tashkun, \& Tennyson]{Rothman2010}
Rothman, L., Gordon, I., Barber, R., Dothe, H., Gamache, R., Goldman, A.,
  Perevalov, V., Tashkun, S., \& Tennyson, J., 2010.
\newblock {HITEMP, the high-temperature molecular spectroscopic database}, {\it
  Journal of Quantitative Spectroscopy and Radiative Transfer\/}, {\bf
  111}(15), 2139--2150.

\bibitem[Sharp \& Burrows(2007)]{07ShBuxx}
Sharp, C.~M. \& Burrows, A., 2007.
\newblock Atomic and molecular opacities for brown dwarf and giant planet
  atmospheres, {\it ApJS\/}, {\bf 168}, 140.

\bibitem[{Sharpe} et~al.(2004){Sharpe}, {Johnson}, {Sams}, {Chu}, {Rhoderick},
  \& {Johnson}]{PNNL-REF}
{Sharpe}, S.~W., {Johnson}, T.~J., {Sams}, R.~L., {Chu}, P.~M., {Rhoderick},
  G.~C., \& {Johnson}, P.~A., 2004.
\newblock {Gas-Phase Databases for Quantitative Infrared Spectroscopy}, {\it
  Applied Spectroscopy\/}, {\bf 58}, 1452--1461.

\bibitem[Shine et~al.(2016)Shine, Campargue, Mondelain, McPheat, Ptashnik, \&
  Weidmann]{16ShCaDi}
Shine, K.~P., Campargue, A., Mondelain, D., McPheat, R.~A., Ptashnik, I.~V., \&
  Weidmann, D., 2016.
\newblock The water vapour continuum in near-infrared windows – current
  understanding and prospects for its inclusion in spectroscopic databases,
  {\it Journal of Molecular Spectroscopy\/}, {\bf 327}, 193--208, New Visions
  of Spectroscopic Databases, Volume II.

\bibitem[{Showman} \& {Guillot}(2002)]{Showman2002}
{Showman}, A.~P. \& {Guillot}, T., 2002.
\newblock {Atmospheric circulation and tides of ``51 Pegasus b-like'' planets},
  {\it \aap\/}, {\bf 385}, 166--180.

\bibitem[Simonova et~al.(2022)Simonova, V.~Ptashnik, Elsey, McPheat, Shine, \&
  Smith]{22SiPtEl}
Simonova, A.~A., V.~Ptashnik, I., Elsey, J., McPheat, R.~A., Shine, K.~P., \&
  Smith, K.~M., 2022.
\newblock Water vapour self-continuum in near-visible ir absorption bands:
  Measurements and semiempirical model of water dimer absorption, {\it
  \jqsrt\/}, {\bf 277}, 107957.

\bibitem[Smith et~al.(1978)Smith, Dube, Gardner, Clough, \&
  Kneizys]{smith1978fascode}
Smith, H., Dube, D., Gardner, M., Clough, S., \& Kneizys, F., 1978.
\newblock Fascode-fast atmospheric signature code (spectral transmittance and
  radiance), Tech. rep., VISIDYNE INC BURLINGTON MA.

\bibitem[{Sung} et~al.(2019){Sung}, {Wishnow}, {Crawford}, {Nemchick},
  {Drouin}, {Toon}, {Yu}, {Payne}, \& {Jiang}]{Sung2019_O2}
{Sung}, K., {Wishnow}, E.~H., {Crawford}, T.~J., {Nemchick}, D., {Drouin},
  B.~J., {Toon}, G.~C., {Yu}, S., {Payne}, V.~H., \& {Jiang}, J.~H., 2019.
\newblock {FTS measurements of O$_{2}$ collision-induced absorption in the
  565-700 nm region using a high pressure gas absorption cell}, {\it \jqsrt\/},
  {\bf 235}, 232--243.

\bibitem[Tan et~al.(2022)Tan, Skinner, Samuels, Hargreaves, Hashemi, \&
  Gordon]{Tan2022}
Tan, Y., Skinner, F.~M., Samuels, S., Hargreaves, R.~J., Hashemi, R., \&
  Gordon, I.~E., 2022.
\newblock {H<sub>2</sub> , He, and CO<sub>2</sub> Pressure-induced Parameters
  for the HITRAN Database. II. Line Lists of CO<sub>2</sub> , N<sub>2</sub>O,
  CO, SO<sub>2</sub> , OH, OCS, H<sub>2</sub>CO, HCN, PH<sub>3</sub>,
  H<sub>2</sub>S, and GeH<sub>4</sub>,}, {\it The Astrophysical Journal
  Supplement Series\/}, {\bf 262}(2), 40.

\bibitem[Tennyson et~al.(2014)Tennyson, Bernath, Campargue, Cs\'asz\'ar,
  Daumont, Gamache, Hodges, Lisak, Naumenko, Rothman, Tran, Zobov, Buldyreva,
  Boone, {De Vizia}, Gianfrani, Hartmann, McPheat, Murray, Ngo, Polyansky, \&
  Weidmann]{jt584}
Tennyson, J., Bernath, P.~F., Campargue, A., Cs\'asz\'ar, A.~G., Daumont, L.,
  Gamache, R.~R., Hodges, J.~T., Lisak, D., Naumenko, O.~V., Rothman, L.~S.,
  Tran, H., Zobov, N.~F., Buldyreva, J., Boone, C.~D., {De Vizia}, M.~D.,
  Gianfrani, L., Hartmann, J.-M., McPheat, R., Murray, J., Ngo, N.~H.,
  Polyansky, O.~L., \& Weidmann, D., 2014.
\newblock {Recommended isolated-line profile for representing high-resolution
  spectroscopic transitions (IUPAC Technical Report)}, {\it Pure Appl.
  Chem.\/}, {\bf 86}, 1931--1943.

\bibitem[Tennyson et~al.(2020)Tennyson, Yurchenko, Al-Refaie, Clark, Chubb,
  Conway, Dewan, Gorman, Hill, Lynas-Gray, Mellor, McKemmish, Owens, Polyansky,
  Semenov, Somogyi, Tinetti, Upadhyay, Waldmann, Wang, Wright, \&
  Yurchenko]{20TeYuAl}
Tennyson, J., Yurchenko, S.~N., Al-Refaie, A.~F., Clark, V.~H., Chubb, K.~L.,
  Conway, E.~K., Dewan, A., Gorman, M.~N., Hill, C., Lynas-Gray, A., Mellor,
  T., McKemmish, L.~K., Owens, A., Polyansky, O.~L., Semenov, M., Somogyi, W.,
  Tinetti, G., Upadhyay, A., Waldmann, I., Wang, Y., Wright, S., \& Yurchenko,
  O.~P., 2020.
\newblock The 2020 release of the exomol database: Molecular line lists for
  exoplanet and other hot atmospheres, {\it Journal of Quantitative
  Spectroscopy and Radiative Transfer\/}, {\bf 255}, 107228.

\bibitem[Thorngren et~al.(2019)Thorngren, Gao, \&
  Fortney]{thorngren2019intrinsic}
Thorngren, D., Gao, P., \& Fortney, J.~J., 2019.
\newblock The intrinsic temperature and radiative--convective boundary depth in
  the atmospheres of hot jupiters, {\it The Astrophysical Journal Letters\/},
  {\bf 884}(1), L6.

\bibitem[Tran et~al.(2013)Tran, Ngo, \& Hartmann]{tran2013efficient}
Tran, H., Ngo, N.~H., \& Hartmann, J.-M., 2013.
\newblock Efficient computation of some speed-dependent isolated line profiles,
  {\it Journal of Quantitative Spectroscopy and Radiative Transfer\/}, {\bf
  129}, 199--203.

\bibitem[Tran et~al.(2014)Tran, Ngo, \& Hartmann]{tran2014erratum}
Tran, H., Ngo, N.~H., \& Hartmann, J.-M., 2014.
\newblock Erratum to “efficient computation of some speed-dependent isolated
  line profiles”[j. quant. spectrosc. radiat. transfer 129 (2013) 199--203],
  {\it Journal of Quantitative Spectroscopy and Radiative Transfer\/}, {\bf
  134}, 104.

\bibitem[Tran et~al.(2022)Tran, {Vander Auwera}, Bertin, Fakhardji, Pirali, \&
  Hartmann]{Tran2022wing}
Tran, H., {Vander Auwera}, J., Bertin, T., Fakhardji, W., Pirali, O., \&
  Hartmann, J.-M., 2022.
\newblock Absorption of methane broadened by carbon dioxide in the 3.3 $\mu$m
  spectral region: From line centers to the far wings, {\it Icarus\/}, {\bf
  384}, 115093.

\bibitem[Underwood et~al.(2016)Underwood, Tennyson, Yurchenko, Huang, Schwenke,
  Lee, Clausen, \& Fateev]{jt635}
Underwood, D.~S., Tennyson, J., Yurchenko, S.~N., Huang, X., Schwenke, D.~W.,
  Lee, T.~J., Clausen, S., \& Fateev, A., 2016.
\newblock {ExoMol line lists XIV: A line list for hot SO$_2$}, {\it Mon. Not.
  R. Astron. Soc.\/}, {\bf 459}, 3890--3899.

\bibitem[Wcis{\l}o et~al.(2016)Wcis{\l}o, Gordon, Tran, Tan, Hu, Campargue,
  Kassi, Romanini, Hill, Kochanov, \& Rothman]{16WcGoTr.H2}
Wcis{\l}o, P., Gordon, I., Tran, H., Tan, Y., Hu, S.-M., Campargue, A., Kassi,
  S., Romanini, D., Hill, C., Kochanov, R., \& Rothman, L., 2016.
\newblock The implementation of non-voigt line profiles in the {HITRAN}
  database: {H$_2$} case study, {\bf 177}, 75 -- 91, XVIIIth Symposium on High
  Resolution Molecular Spectroscopy (HighRus-2015), Tomsk, Russia.

\bibitem[Yurchenko et~al.(2017)Yurchenko, Amundsen, Tennyson, \&
  Waldmann]{jt698}
Yurchenko, S.~N., Amundsen, D.~S., Tennyson, J., \& Waldmann, I.~P., 2017.
\newblock {A hybrid line list for CH$_4$ and hot methane continuum}, {\it
  A\&A\/}, {\bf 605}, A95.

\bibitem[Yurchenko et~al.(2018)Yurchenko, Al-Refaie, \& Tennyson]{ExoCross}
Yurchenko, S.~N., Al-Refaie, A.~F., \& Tennyson, J., 2018.
\newblock {EXOCROSS}: a general program for generating spectra from molecular
  line lists, {\it A\&A\/}, {\bf 614}, A131.

\bibitem[Zhang et~al.(2023)Zhang, Tennyson, \& Yurchenko]{PyExoCross}
Zhang, J., Tennyson, J., \& Yurchenko, S.~N., 2023.
\newblock {PyExoCross: a Python program for generating spectra from molecular
  line lists}, {\it RASTI\/}, (in preparation).

\bibitem[Zhang et~al.(2020)Zhang, Chachan, Kempton, Knutson,
  et~al.]{zhang2020platon}
Zhang, M., Chachan, Y., Kempton, E. M.-R., Knutson, H.~A., et~al., 2020.
\newblock Platon ii: New capabilities and a comprehensive retrieval on hd
  189733b transit and eclipse data, {\it The Astrophysical Journal\/}, {\bf
  899}(1), 27.

\end{thebibliography}







\bsp	
\label{lastpage}
\end{document}